\documentclass[fleqn,usenatbib]{mnras}

\usepackage{newtxtext,newtxmath}

\usepackage[T1]{fontenc}

\DeclareRobustCommand{\VAN}[3]{#2}
\let\VANthebibliography\thebibliography
\def\thebibliography{\DeclareRobustCommand{\VAN}[3]{##3}\VANthebibliography}

\usepackage{graphicx}	
\usepackage{amsmath}	

\usepackage[caption=false]{subfig}
\usepackage{pythonhighlight}
\usepackage[utf8]{inputenc} 

\title[Verifying methods tracing galaxy disturbances]{Comparison and verification methods to trace interaction-driven disturbances in galaxies}

\author[H. Lyu et al.]{
Haotian Lyu$^{1}$,
Sarah Brough$^{1}$,\thanks{E-mail: s.brough@unsw.edu.au}
Aman Khalid$^{1}$,
Alice Desmons$^{1}$,
and Elizaveta Sazonova$^{2,3}$
\\
$^{1}$School of Physics, University of New South Wales, NSW 2052, Australia\\
$^{2}$Waterloo Centre for Astrophysics, University of Waterloo, Waterloo, ON N2L 3G1, Canada\\
$^{3}$Department of Physics and Astronomy, University of Waterloo, Waterloo, ON N2L 3G1, Canada
}

\date{Accepted XXX. Received YYY; in original form ZZZ}

\pubyear{\the\year{}}

\begin{document}
\label{firstpage}
\pagerange{\pageref{firstpage}--\pageref{lastpage}}
\maketitle

\begin{abstract}
Low surface brightness tidal debris around galaxies, such as tails, streams, and shells, together with other interaction-driven morphological disturbances, serve as valuable indicators of past or ongoing galaxy mergers. With the growing data volume from surveys like the Vera C. Rubin Observatory's Legacy Survey of Space and Time (LSST), automated detection methods are essential. This paper evaluates the performance of two automated methods, a Self-Supervised Learning (SSL) model and the Concentration-Asymmetry-Smoothness (CAS) parameter method, in tracing interaction-driven disturbances and merger signatures, with visual classification used as the benchmark. Visual classification yields a high-confidence disturbance fraction of $25.1 \pm 1.5\%$ in our sample and serves as the reference standard for assessing the completeness and precision of the automated approaches. Visual classification is affected by galaxy distance and image resolution, which limit the detectability of faint low surface brightness structures. The SSL model achieves high recall ($0.86 \pm 0.04$) and low contamination (0.2) by retraining only its linear classifier on a small labelled dataset, making it suitable for identifying a broad set of disturbed systems, including faint tidal debris and other interaction-driven morphological disturbances, thereby providing a more complete census of merger-related features. The CAS method, using the traditional threshold $A > 0.35$, shows higher precision (0.77) but lower recall (0.20), indicating a conservative approach that captures cleaner but less complete samples. Visual classification and the SSL model show a significant positive correlation between stellar mass and disturbance fraction, while the CAS method exhibits a much weaker trend.

\end{abstract}




\section{Introduction} \label{sec:1}
A key prediction of the Lambda Cold Dark Matter ($\Lambda$CDM) cosmological model is the hierarchical formation of cosmological structures (e.g. \citealt{dalal_2008_halo}). According to this paradigm, the formation of structures, such as galaxies and galaxy clusters, occurs from the bottom up. This means that larger structures are formed through the continuous merging of smaller structures (e.g. \citealt{lacey_1993_merger}, \citealt{dalal_2008_halo}, \citealt{10.1046/j.1365-8711.2000.03879.x}, \citealt{martin_2018_the}). During the merging process, galaxy morphologies are significantly altered. Gravitational interactions between merging galaxies will pull stars out of their bound orbits \citep{1972ApJ...178..623T}. This results in the formation of diffuse, irregular regions known as tidal features. Due to the range of angles, speeds, and angular momentum of galaxy mergers, as well as the different properties of the parent galaxies, tidal features exhibit diverse morphologies and colours. These characteristics provide valuable insights into the dynamics of the merging process. Specifically, the presence and properties of tidal features provide constraints on merger histories, while their colours can offer information about the stellar populations of the parent galaxies involved in these interactions (e.g. \citealt{kadofong_2018_tidal}, \citealt{pop_2018_formation}, \citealt{karademir_2019_the}). Additionally, the morphology of the tidal features conveys information about the initial angular momentum of the merger (e.g. \citealt{10.1093/mnras/stv2035}). While tails, streams, and shells represent specific low-surface-brightness tidal debris structures, galaxy interactions and mergers can also give rise to broader morphological disturbances, such as asymmetric outer haloes and double nuclei. In this work, we treat both as useful tracers of interaction-driven galaxy evolution.

In the past, the identification and classification of tidal features and other merger-related morphological disturbances relied on visual classification (e.g. \citealt{tal_2009_the}, \citealt{atkinson_2013_faint}, \citealt{martin_2022_preparing}). The low surface brightness of many tidal debris structures made them difficult to detect in previous wide-field optical astronomical surveys, resulting in a lack of sufficiently large samples. Consequently, visual classification was deemed sufficient for such classification tasks. However, with the advent of the Vera C. Rubin Observatory's Legacy Survey of Space and Time (LSST; \citealt{ivezi_2019_lsst}), the new images will be sensitive enough to detect tidal features, generating a vast amount of data for millions of galaxies (e.g. \citealt{robertson_2019_galaxy}). In addition, the Euclid mission \citep{2025A&A...697A...1E} is already delivering high-resolution imaging over thousands of square degrees, providing large samples of galaxies with resolved tidal debris (e.g. \citealt{2025A&A...700A.104U}). Observational studies of nearby galaxies at comparable surface brightness depths, such as \citet{Sola2022} and \citet{Sola2025}, have already demonstrated that tidal features are detectable in a significant fraction of massive galaxies at limits of $\sim$29 mag arcsec$^{-2}$, providing an empirical baseline ahead of the LSST era. Together, these surveys will generate imaging data on a scale for which purely visual classification will no longer be feasible, underscoring the urgent need for robust automated tools to identify interaction-driven disturbances and merger signatures.

In recent years, there has been a growing use of machine learning (ML) for identifying tidal debris and other interaction-driven morphological disturbances, largely to solve the limitations of visual classification. Supervised learning methods such as convolutional neural networks (CNNs) have proven effective in both galaxy morphological classification (e.g. \citealt{walmsley_2019_galaxy}, \citealt{cheng_2020_optimizing}) and faint tidal feature identification (e.g. \citealt{walmsley_2018_identification}, \citealt{bickley_2021_convolutional}, \citealt{2023MNRAS.521.3861D}, \citealt{2023A&A...679A.142O}), offering greater efficiency and reproducibility across large datasets. However, these approaches often rely on large, high-quality labelled datasets, which are rarely available for faint tidal features. As a result, some studies (e.g. \citealt{bickley_2021_convolutional}, \citealt{hdomnguezsnchez_2023_identification}, \citealt{10.1093/mnras/stae2246}, \cite{refId0}) have resorted to using simulated images to compensate for the lack of real labelled data. Unlike supervised learning models, unsupervised learning does not require labelled training samples but classifies images by self-summarising features (e.g. \citealt{hocking_2017_an}, \citealt{martin_2020_galaxy}). However, unsupervised learning can only cluster and classify data without matching them to labels. As a middle ground, self-supervised learning (SSL) uses unlabelled images to train the self-supervised model, which converts the images into meaningful low-dimensional representations and then uses a smaller labelled data set to train a linear classifier to provide the final classifications. The self-supervised model is used as the encoder of the linear classifier. Notably, \citet{desmons_2023_detecting} demonstrated an SSL-based method that achieves both high completeness and low contamination in detecting tidal features in ultradeep layer images from Hyper Suprime-Cam Subaru Strategic Program (HSC-SSP; \citealt{10.1093/pasj/psz103}).

Another widely used automated approach for tracing galaxy interactions and mergers is the CAS (concentration, asymmetry, smoothness) parameter method. This non-parametric technique quantifies the morphological structure of galaxies from their light distributions \citep{conselice_2008_the}, based on the idea that the light distribution of galaxies provides insights into their past and present formation modes \citep{conselice_2003_the}. Among the three CAS parameters, asymmetry is considered the main indicator of galaxy mergers. A higher value of asymmetry means an asymmetric light distribution, which is usually found in spiral galaxies and in systems undergoing interactions \citep{conselice_2008_the}. Unlike methods aimed at identifying specific tidal debris structures, CAS primarily traces global morphological disturbance through the galaxy light distribution. The other two parameters also provide insights into merging events. For example, the comparison between asymmetry and smoothness can also be used as a reference. Usually, an asymmetry parameter $A > 0.35$ and a higher value of asymmetry than for smoothness are considered criteria for identifying merging galaxies.

In this paper, we use visual classification as a baseline to evaluate two automated approaches, the SSL model and the CAS parameter method, for tracing interaction-driven disturbances and merger signatures. We apply these methods to a new set of mock images generated using the same methods as \citet{khalid_2023_characterising}, and systematically assess their performance in recovering both low-surface-brightness tidal features and broader disturbed morphologies. By comparing the automated results with those from visual classification, we highlight the advantages and limitations of each approach. In Section \ref{sec:2}, we detail the data sources and processing procedures, and introduce the two automated recognition methods used in this study, including their respective data augmentation strategies. In Section \ref{sec:3}, we present the results obtained from the visual classification and two automated classification methods. Section \ref{sec:4} compares the results of the different automated methods for tracing interaction-driven disturbances and further contrasts them with findings in previous literature. Building on these comparisons and findings, we present our conclusions in Section \ref{sec:5}. 

Throughout this paper, we assume a flat $\Lambda$CDM cosmology with parameters from \textit{Planck} 2015 \citep{ade_2016_planck2015}: $H_0 = 67.8$ km s$^{-1}$ Mpc$^{-1}$, $\Omega_m = 0.308$, and $\Omega_\Lambda = 0.692$.

\section{Data and Methods} \label{sec:2}
\subsection{Data Sources} \label{sec:2.1}
We use deep mock images generated from cosmological simulations to evaluate the performance of the SSL model developed by \citet{desmons_2023_detecting} and the CAS parameters in tracing interaction-driven disturbances, including low surface brightness tidal features. Mock images are employed here because they can be generated to closely resemble the upcoming LSST data. This similarity provides a valuable testing ground for applying automated methods for identifying interaction-driven disturbances and merger signatures in future LSST imaging. The initial dataset was produced by \citet{khalid_2023_characterising} using the \textsc{IllustrisTNG-100} simulation \citep{10.1093/mnras/stx3304}. The simulation has a stellar mass resolution of \( m_{\star} = 1.1 \times 10^{6} \, M_{\odot} \) and dark matter particle resolution of \( m_{DM} = 7.5 \times 10^6 M_{\odot}\). This leads to a sample of 1826 galaxies with stellar masses ranging from \(3.16 \times 10^9\,M_{\odot}\) to \(7.0 \times 10^{11}\,M_{\odot}\) (mean \(2.7 \pm 0.1 \times 10^{10} \,M_{\odot}\)). Stellar masses are computed by summing the masses of all stellar particles within a 30 kpc aperture around the galaxy's centre of potential. 

\begin{figure*}
    \centering

    \subfloat[]{%
        \includegraphics[height=3.8cm]{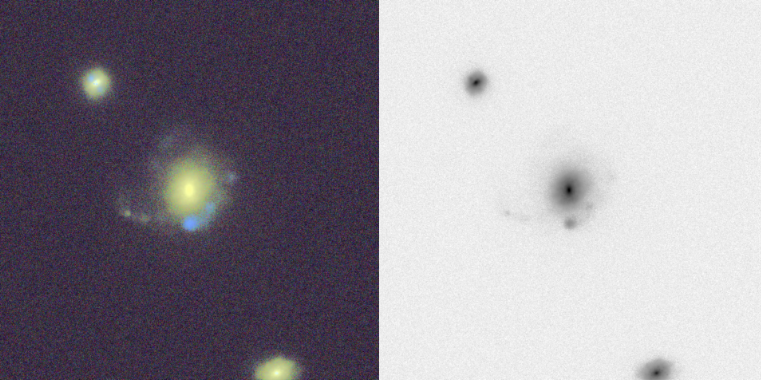}
        \label{fig:1a}
    }\subfloat[]{%
       \includegraphics[height=3.8cm]{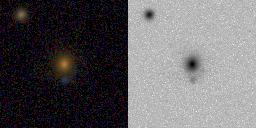}
       \label{fig:1b}
    } \\
    \subfloat[]{%
       \includegraphics[height=3.8cm]{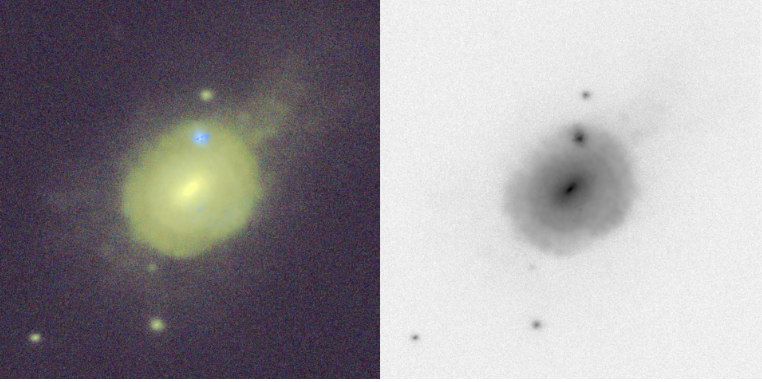}
       \label{fig:1c}
    } \subfloat[]{%
       \includegraphics[height=3.8cm]{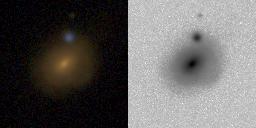}
       \label{fig:1d}
    }

\caption{Example galaxies in the original and re-made mock image sample. Panels (a) and (b) show the same galaxy, and Panels (c) and (d) show another galaxy. Panel (a) and Panel (c) are from \citet{khalid_2023_characterising}'s original mock images, with a size of 2400 $\times$ 2400 pixels and placed at a distance of $z \sim 0.025$. Panel (b) and Panel (d) are the re-made mock images, with a size of 128 $\times$ 128 pixels and placed at a distance of $z \sim 0.2$. The left panels of each figure show the \textit{gri}-coloured images, while the right panels show the corresponding \textit{g}-band grayscaled images.}
\label{fig:1}
\end{figure*}

To construct these mock images, \citet{khalid_2023_characterising} selected all the stellar particles within a 1 Mpc side length cube of the galaxy of interest by using the FOF (Friends-of-Friends) and SubFind catalogues of halos \citep{2001MNRAS.328..726S, 2009MNRAS.399..497D}. The spectral energy distributions (SEDs) to each particle are assigned using the stellar population models from \citet{10.1046/j.1365-8711.2003.06897.x}, interpolated by age and metallicity. They assume a Chabrier initial mass function \citep{2003PASP..115..763C}, and assign \textit{grizy} flux values to each particle, accounting for the galaxy redshift and the LSST bandpass functions \citep{olivier_2008_optical}. Pixels containing only a few stellar particles were smoothed by subdividing the stellar particles into smaller mass particles, with their positions normally distributed around the original particle. The 3D cubes were projected to 2D, convolved with the HSC-SSP point spread function measured by \citet{montes_2021_the} and rescaled to match an LSST-like spatial resolution of \(0.2^{\prime\prime}\,\mathrm{pixel}^{-1}\). The Gaussian noise was added to emulate the expected 10-year LSST surface brightness limits: $\mu_g \sim 30.3$, $\mu_r \sim 30.3$, $\mu_i \sim 29.7$, $\mu_z \sim 28.4$, and $\mu_y \sim 28.1$ mag arcsec$^{-2}$ (\citealt{yoachim_2024_surface}; 3$\sigma$, $10'' \times 10''$).

\citet{khalid_2023_characterising} developed mock images at a redshift of \(z \sim 0.025\) ($\sim 105\text{ Mpc}$) with a size of 2400 pixels $\times$ 2400 pixels for the purpose of visual classification. As one of the purposes of this paper is to test the \citet{desmons_2023_detecting} model, we re-made the mock images to be consistent with those used in that paper (their median $z = 0.15$). This choice was motivated by the fact that the semi-supervised model of  \citet{desmons_2023_detecting} was designed and trained to detect tidal features around LSST-like galaxy populations, which will be visible to redshifts around $z \sim 0.15$ (e.g. \citealt{martin_2022_preparing}). The very nearby systems in \citet{khalid_2023_characterising} ($z \sim 0.025$) are not representative of the bulk of the LSST sample, and therefore re-making the mock images at LSST-like redshifts provides a more appropriate test of the model’s performance. The re-made mock images were positioned at $z \sim 0.2$ ($\sim 843$ Mpc), spanning the range $0.02 < z < 0.29$. This redshift distribution was generated by sampling from a Gaussian distribution centred at $z = 0.15$, which equals the median redshift of the \citet{desmons_2023_detecting}'s sample. The standard deviation of the re-made redshift distribution, 
$\sigma_z = 0.04$, was chosen to match that of \citet{desmons_2023_detecting}'s sample, ensuring a consistent redshift distribution.

The re-made mock images were initially generated with a size of 278 X 278 pixels, which is $56^{\prime\prime} \times 56^{\prime\prime}$ at the LSST pixel scale of $0.2^{\prime\prime}$ per pixel. This image size was chosen based on \citet{desmons_2023_detecting}, who demonstrated that an image size of $60^{\prime\prime} \times 60^{\prime\prime}$ is more than sufficient for capturing tidal features over this redshift range. The slightly larger size provides a margin of flexibility, accounting for potential numerical uncertainties introduced by the redshift-based conversion from kiloparsecs to arcseconds during image generation. The re-made mock images were then cropped to \(256 \times 256\) pixels and rebinned to a final size of \(128 \times 128\) pixels (corresponding to a resolution of \(0.43^{\prime\prime}\,\mathrm{pixel}^{-1}\)) to match the \(128 \times 128\) pixels input requirements of the SSL model from \citet{desmons_2023_detecting}.
Additionally, while the original mock images from \citet{khalid_2023_characterising} used only the \emph{gri}-bands, we generated the full set of \emph{grizy} mock images following the same particle–SED and filter-convolution procedure described above, rather than adjusting the original images. This ensured that all five HSC-SSP bands (\emph{grizy}) were consistently produced from the underlying particle data to match the input format required by the SSL model.

Figure \ref{fig:1} compares the original and re-made mock images of two galaxies. In Panel (a), the original mock image clearly shows a stream/tail, but in the re-made version (Panel b), this feature becomes difficult to identify due to the effects of increased distance. A similar change is observed between Panel (c) and Panel (d), where a structure identified as a shell in the original image appears more like an asymmetric halo in the re-made image. These differences arise primarily from surface brightness dimming caused by greater distance, which reduces the visibility of faint features.

To facilitate comparison with the machine learning classification results, we visually reclassified the re-made mock image sample. The details of the visual classification process are provided in the next section.

\subsection{Visual Classification} \label{sec:2.2}
\begin{figure*}
    \centering
    \subfloat[]{
        \includegraphics[width=0.8\textwidth]{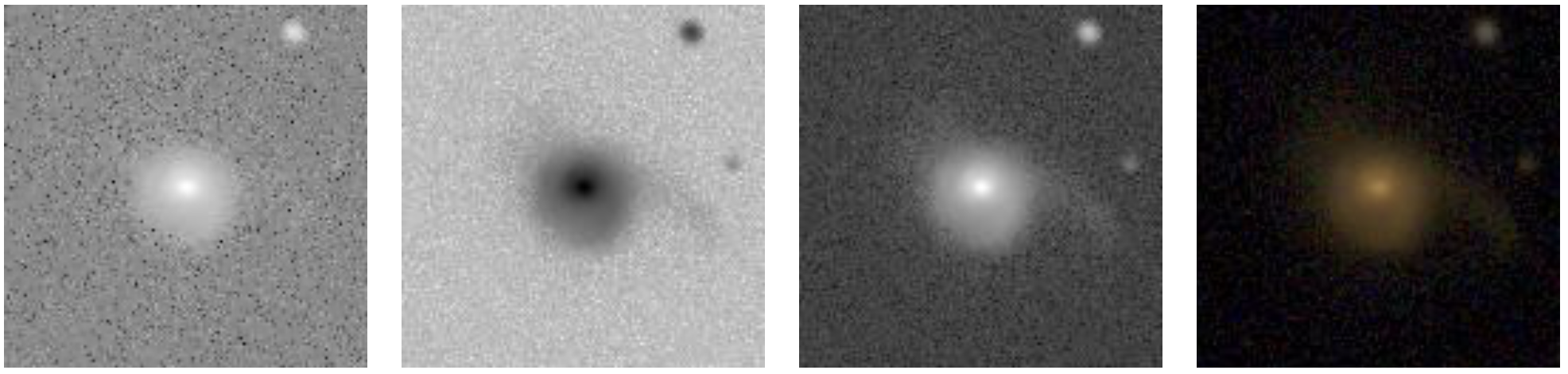}
        \label{fig:2a}
    }
    
    \vspace{-6pt}

    \subfloat[]{
        \includegraphics[width=0.8\textwidth]{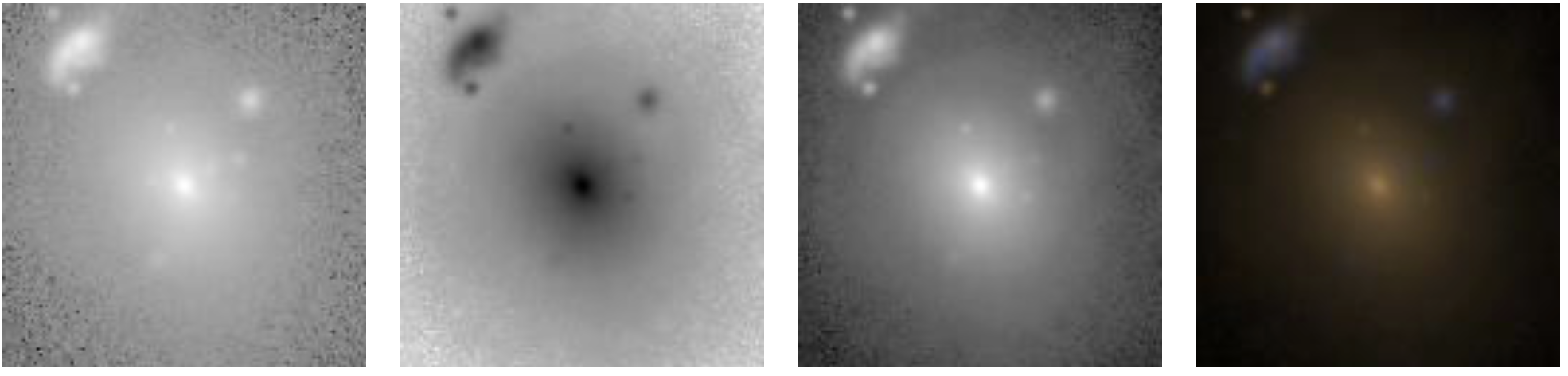}
        \label{fig:2b}
    }

    \vspace{-6pt}

    \subfloat[]{
        \includegraphics[width=0.8\textwidth]{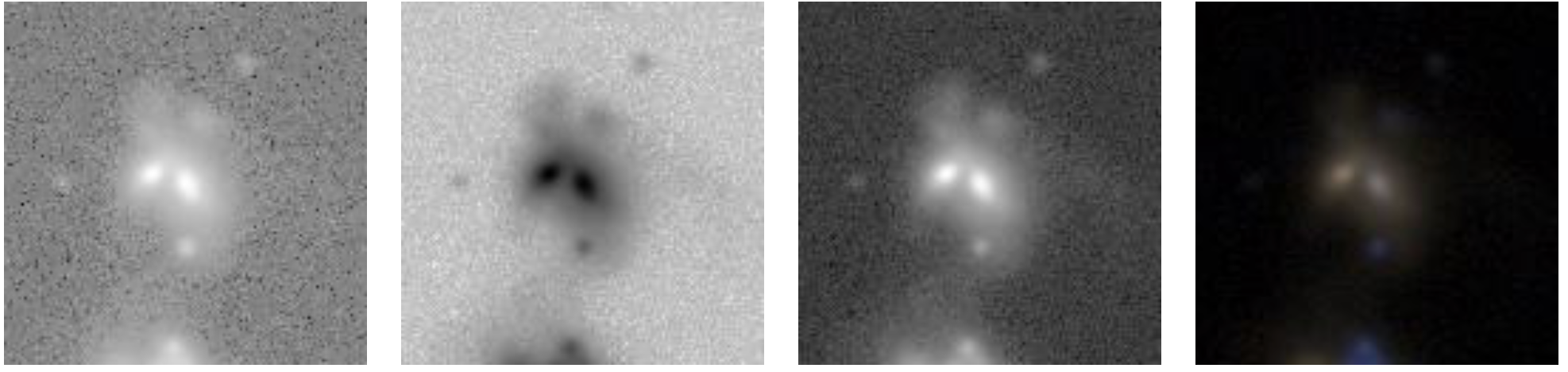}
        \label{fig:2c}
    }

    \caption{Example galaxies in the remade mock-image sample. The four images from left to right in each row show \textit{g}-band grayscale, 5-band grayscale, 5-band inverse grayscale, and \textit{gri}-colour, respectively. Row (a) shows a galaxy classified as containing a stream/tail and an asymmetric halo. Row (b) is classified as showing a shell, and row (c) is classified as showing a double nucleus and an asymmetric halo (confidence 3). All displayed features are classified at confidence level 3. The description of the confidence levels is given in Table \ref{table:1}.}

    \label{fig.2}
\end{figure*}

The visual classification of the re-made mock images is based on the classification criteria applied in \citet{khalid_2023_characterising} and \citet{desmons_2023_galaxy}, which follows a simplified version of the \citet{blek_2020_census} scheme. \citet{khalid_2023_characterising} and \citet{desmons_2023_detecting} merged the categories of streams and tails since they are easily mistaken for one another at the spatial resolution of LSST.

We visually classify interaction-driven disturbances into four categories. Streams/tails and shells are treated as specific tidal features, while asymmetric stellar haloes and double nuclei are treated as broader disturbance signatures associated with galaxy interactions and mergers:

\begin{itemize}
  \item \textbf{Streams/Tails:} Prominent, elongated structures orbiting or expelled from the host galaxy. Or these structures can also originate from tidally stripped companion galaxies.
  \item \textbf{Shells}: Concentric radial arcs or ring-like structures around a galaxy.
  \item \textbf{Asymmetric Stellar Halos}: Diffuse features in the outskirts of the host galaxy, lacking well-defined structures like stellar streams or tails. These features are treated here as broader disturbance signatures rather than as specific tidal debris structures.
  \item \textbf{Double nuclei}: Two clearly separated luminous cores or galaxies in close proximity within a common system, consistent with an ongoing merger. Given that the classification is based on photometric imaging rather than spectroscopic confirmation, additional evidence in the form of faint connecting or surrounding disturbance between or around the two components is required to reduce the likelihood of chance projections. Such evidence need not take the form of prominent tails, streams, or shells.
\end{itemize}

These four categories trace related but non-identical manifestations of galaxy interactions and mergers. Streams/tails and shells are treated as specific tidal debris structures, whereas asymmetric stellar haloes and double nuclei are broader disturbance signatures. In particular, the visibility of a double nucleus does not require equally prominent low surface brightness tidal debris, because the nuclear structure and the outer debris can have different detectability and visibility timescales in photometric imaging. As a result, a system may be assigned a high confidence in the double-nucleus category even when tails, streams, or shells are weak or not clearly detectable.

To comprehensively assess the morphologies of low surface brightness galaxies and classify them accurately, we employed four imaging modalities for each galaxy: the individual \textit{grizy} bands, grayscale and inverse grayscale images combining all five bands, and \textit{gri}-colour images for the visual classification process. Composite 5-band grayscale and inverse grayscale images were created by combining the flux from all five bands per pixel, which improves the contrast between tidal features and the background. Combined \textit{gri} images are constructed using the \textit{g}, \textit{r}, and \textit{i} bands.

Figure \ref{fig.2} presents examples of the remade mock images using the different visualisation modalities. For the single-band representation, we show only the \textit{g}-band grayscale image, while the remaining panels display the 5-band grayscale, 5-band inverse grayscale, and \textit{gri}-colour composites. In practice, however, the full classification process utilises all five bands to identify both low surface brightness tidal debris and broader interaction-driven disturbance signatures. Different imaging modalities reveal distinct structural details. For instance, in Figure \ref{fig.2} panel (a), the stream/tail is difficult to discern in the \textit{g}-band grayscale image, but it becomes clearly visible in the 5-band grayscale and inverse grayscale images.

All images used for visual classification are displayed with a logarithmic flux scaling. This choice enhances the visibility of low surface brightness structures and produces a contrast that is more natural for human inspection, since the logarithmic transformation produces a brightness scaling comparable to that of magnitude units. This scaling was consistently applied across all images to ensure classification uniformity.

\begin{table}
\caption{Descriptions of the visual-classification confidence levels \citep{khalid_2023_characterising}.}
\centering
\begin{tabular}{c | p{6cm}} 
 \hline
 Confidence & Description \\ [0.5ex] 
 level &  \\ [0.5ex] 
 \hline\hline
 0 & No interaction-driven disturbance detected.  \\ [1ex] 
 \hline
 1 & Hint of a disturbance detected, but classification remains uncertain. \\[1ex] 
 \hline 
 2 & Moderate confidence that a disturbance is present and/or that its morphology has been correctly identified. \\[1ex] 
 \hline 
 3 & High confidence that the disturbance is present and that its morphology is clearly identifiable. \\[1ex] 
 \hline
\end{tabular}
\label{table:1}
\end{table}

To evaluate the visual classification of interaction-driven disturbances, we use the same confidence system as \citet{khalid_2023_characterising} to quantify the likelihood that a given disturbance signature is present. Table \ref{table:1} shows the classification confidence levels together with their corresponding descriptions. The highest confidence level across the four categories was adopted as the final confidence level for each galaxy. Because the categories trace related but non-identical disturbance signatures, this final confidence is used as an operational summary of the strongest visually identified interaction-driven disturbance in each system.

\subsection{Self-Supervised Learning Model} \label{sec:2.3}
We evaluate the performance of the self-supervised learning (SSL) model presented in \citet{desmons_2023_detecting} in classifying interaction-driven disturbances in the remade mock image sample. To facilitate this evaluation, we used the results from visual classification as the ground truth for the dataset. Based on the confidence levels of the visual classifications (Table \ref{table:1}), we assign a final confidence level to each mock image by taking the highest confidence level among the four visually identified disturbance categories. Galaxies with a confidence level of 2 or higher are assigned positive labels, while those with confidence levels of 0 or 1 are given negative labels.

\subsubsection{Model Structure} \label{sec:2.3.1}
\citet{desmons_2023_detecting}'s SSL model consists of a convolutional encoder trained using contrastive learning to extract low-dimensional representations of galaxy images, followed by a linear classifier that distinguishes galaxies with and without visually identified interaction-driven disturbances.

The encoder is a ResNet-20 convolutional neural network \citep{7780459}, which transforms galaxy images into a 128-dimensional feature space. Instead of relying on labelled data, the model is trained through Nearest Neighbour Contrastive Learning of visual Representations (NNCLR; \citealt{dwibedi_2021_with}), where augmented versions of the same image form positive pairs, while different images act as negative pairs. 

The training objective is to minimise the distance between positive pairs while maximising the distance between negative pairs in feature space. Unlike standard contrastive methods, NNCLR enhances feature representations by dynamically selecting positive samples from the nearest neighbours stored in a memory queue, improving the encoder's ability to generalise across different galaxy morphologies.  During training, each input image undergoes a series of augmentations (see Section \ref{sec:2.3.2}) which ensure that the model learns invariant features, making it robust to different types of variations. The encoder is trained using an Adam optimiser \citep{kingma2017adammethodstochasticoptimization} with a learning rate of 0.001 and a batch size of 512. The contrastive loss function follows the InfoNCE formulation \citep{oord2019representationlearningcontrastivepredictive}, optimising the embeddings to cluster similar galaxies together in representation space. The original encoder in \citet{desmons_2023_detecting} is trained over 25 epochs using a dataset of ~44,000 unlabelled galaxies from the Ultradeep layer of the Hyper Suprime-Cam Subaru Strategic Program (HSC-SSP; \citealt{10.1093/pasj/psz103}).

After the encoder is pre-trained, a linear classifier is applied to the learned feature representations to perform binary classification. This classifier consists of a single fully connected layer with sigmoid activation, which assigns a probability score indicating whether a galaxy exhibits an interaction-driven disturbance according to the adopted binary labelling scheme. Unlike fully supervised models that require large labelled datasets, this original linear classifier in \citet{desmons_2023_detecting} is trained using only 600 labelled galaxies (300 with tidal features and 300 without), using binary cross-entropy loss and an Adam optimiser. This self-supervised approach enables efficient automated identification of disturbed galaxies with minimal reliance on labelled data, making it well-suited for large-scale galaxy surveys.

\subsubsection{Image Pre-processing and Augmentations} \label{sec:2.3.2}
In order to ensure consistency with the data input to the trained model, we use the same image pre-processing methods as \citet{desmons_2023_detecting} to normalise the mock images. However, unlike their approach with their larger HSC-SSP sample, which only calculates the standard deviation of a subsample of the first 1,000 galaxies, we use the median absolute deviation to calculate the standard deviation, $\sigma$, of the pixel values in each band (\textit{g, r, i, z, y}) of all 1,826 mock images directly. This adjustment accounts for the smaller sample size in our dataset. Then, each band of the mock image sample is normalised by dividing by 3 $\times$ the corresponding $\sigma$ and then applying an inverse hyperbolic sine function, consistent with \citet{desmons_2023_detecting}.

As we mentioned in Section \ref{sec:2.3.1}, the encoder of the SSL model uses the augmented version of the original images to learn which augmentation transformations preserve the low-dimensional representations of the original images. Based on the augmentations from \citet{desmons_2023_detecting} including \textbf{Flip}, \textbf{Gaussian Noise}, and \textbf{Jitter and Crop}, we also added two more augmentations (\textbf{Rotation} and \textbf{Zooming}) to improve the performance of the model and finally include the following:

\begin{itemize}
    \item \textbf{Flip}: The images are randomly flipped across the $x$ or $y$ axis.
    \item \textbf{Gaussian Noise}: We multiply a scalar generated from the uniform distribution between 1 to 3 with the median absolute deviation $\sigma_{\text{pixel count}}$ to get a per-channel noise $\sigma_c$. Then, a Gaussian noise is generated by using $\sigma_c$ to multiply a normal distribution from 0 to 1. Finally, we added this Gaussian noise to each image.
    \item \textbf{Jitter and Crop}: The mock images with the size of 128 $\times$ 128 pixels are cropped to the central 109 $\times$ 109 pixels and continue to be cropped to 96 $\times$ 96 pixels with random centres, which leads to changes to the image centre. The galaxy is originally located at the centre of the image, but after cropping, it is offset from the centre of the image by a range of $(-13, 13)$ in both $x$ and $y$ axes.
    \item \textbf{Rotation}: The images are randomly rotated, and the rotational angle is in the range $[-0.4 \pi, 0.4 \pi ]$.
    \item \textbf{Zooming}:  We randomly apply zooming to the images, scaling them in or out by up to $20\%$.
\end{itemize}

\subsubsection{Model Evaluation} \label{sec:2.3.3}
For the automatic classification of interaction-driven disturbances, our main concern is the prediction performance for positive cases (images classified as disturbed under the adopted binary labelling scheme). Therefore, we use the True Positive Rate (Recall or Completeness) to assess the model's performance in correctly identifying disturbed systems and the False Positive Rate (Contamination) to measure the proportion of incorrect positive predictions.

True Positive Rate (TPR) and False Positive Rate (FPR) all range from 0 to 1, and are defined as:
\begin{equation}
\begin{aligned}
    TPR &= \frac{TP}{TP + FN} \\
    FPR &= \frac{FP}{FP + TN}
\end{aligned}
\end{equation}
where TP, TN, FP, and FN refer to True Positive, True Negative, False Positive, and False Negative, respectively. The Receiver Operating Characteristic (ROC) curve provides a comprehensive assessment of the classifier's performance by plotting the TPR against the FPR across various threshold settings. The SSL model outputs continuous probabilities between 0 and 1, while the image labels are binary (0 or 1). Thus, an appropriate threshold is required to convert these probabilities into binary classifications (disturbed versus non-disturbed). The ROC curve facilitates this by identifying the threshold that balances a high TPR with a low FPR, optimising the trade-off between identifying disturbed systems and minimising false positives. The area under the ROC curve (AUC) quantifies the model's discriminative ability: an AUC of 0.5 corresponds to random classification, while values closer to 1 indicate stronger performance, with 1 representing perfect classification \citep{narkhede_2018_understanding}. Following \citet{desmons_2023_detecting}, we adopt the TPR at an FPR of 0.2 as the final measure of the model's completeness, selecting the corresponding threshold to convert continuous outputs into binary predictions.

In addition to the TPR, FPR, and ROC AUC, we incorporate additional metrics to further evaluate and compare the model's performance on whole datasets, specifically selecting Precision and F1 Score for a more comprehensive assessment, they are defined as:
\begin{equation}
\begin{aligned}
    \text{Precision} &= \frac{TP}{TP + FP} \\
    \text{F1 score} &= 2 \cdot \frac{\text{Precision} \cdot \text{Recall}}{\text{Precision} + \text{Recall}}
\end{aligned}
\end{equation}
Precision measures the proportion of true positives out of all predicted positives. This captures different aspects of the model's performance to recall (TPR): recall indicates the model's ability to identify all relevant positive instances, while precision reflects the accuracy of its positive predictions. The F1 Score, as the harmonic mean of precision and recall, provides a more comprehensive measure of the model's performance.

\subsection{CAS Parameters} \label{sec:2.4}
The CAS parameters provide a non-parametric way to trace galaxy interactions and mergers through global morphological disturbance. Following \citet{conselice_2003_the}, we adopt the standard merger-selection criteria $A > 0.35$ and $A > S$, and therefore focus here on the asymmetry ($A$) and smoothness ($S$) parameters. For our 5-band mock images, we enhance the image contrast and improve the signal-to-noise ratio by combining the 5 bands into a combined image on which to calculate the CAS parameters. 

\subsubsection{Mask Generation} \label{sec:2.4.1}
Mask generation is a critical part of our analysis pipeline, since it defines the spatial region of the target galaxy, which is the central galaxy in each mock image. This mask is used during the calculation of CAS parameters to ensure that measurements are restricted to the relevant region of the image. Our mask-generation procedure is implemented in \pyth{Python} and follows a modular, parameterised design. The \pyth{photutils} package \citep{larry_bradley_2024_12585239} provides useful functions for estimating the 2D background noise and image segmentation, which are used to detect sources and deblend them to create masks.

\begin{figure}
    \centering
    \subfloat[No smoothing]{%
        \includegraphics[width=0.23\textwidth]{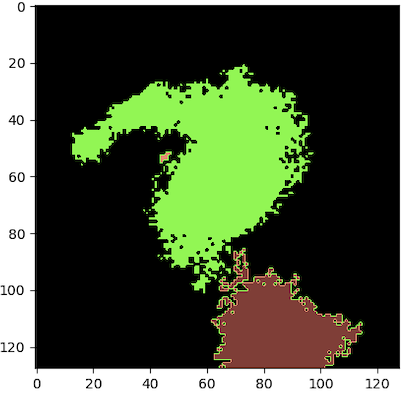}
        \label{fig:3a}
    }
    \subfloat[Smoothing]{%
        \includegraphics[width=0.23\textwidth]{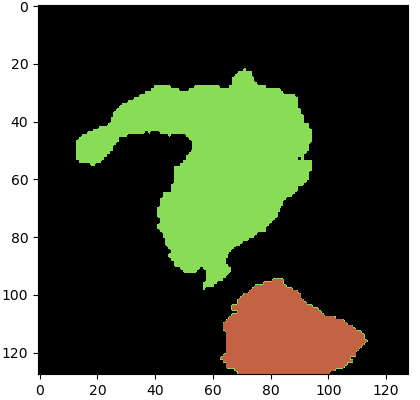}
        \label{fig:3b}
    }
    \caption{Comparison of segmentation maps for the same image generated without and with image smoothing. Panel (a) shows the segmentation map produced without smoothing, while panel (b) displays the segmentation map generated after applying smoothing.}
    \label{fig:3}
\end{figure}

We estimate the 2D background using the \pyth{MedianBackground} estimator within the \pyth{Background2D} function from \pyth{photutils.background}, adopting a filter size of (3, 3). This background model is used solely to generate masks for source detection and is not subtracted from the mock images at any stage. As the mock images are constructed with a low and spatially uniform background, its impact is minimal. To separate the light source from the background for the masking, we apply \pyth{detect_threshold} from \pyth{photutils.segmentation}, setting the threshold at $1.1\sigma$ above the estimated 2D background level. The choice of background filter size, thresholds, and Gaussian smoothing parameters follows the method outlined by \citet{10.1093/mnras/stac3119} for masking images of similar depth. Finally, we smooth the galaxy images by convolving them with a Gaussian filter of $\sigma = 5$ pixels, implemented via the \pyth{Gaussian2DKernel} function from \pyth{astropy} \citep{astropy:2022}. Figure \ref{fig:3} compares segmentation maps of the same image with and without smoothing. Smoothing is essential since the light intensity at the galaxy's edges gradually diminishes and approaches the background noise level. Without it, noise fluctuations cause irregular edges, holes, and scattered pixels.

\begin{figure}
    \centering
    \includegraphics[width=0.48\textwidth]{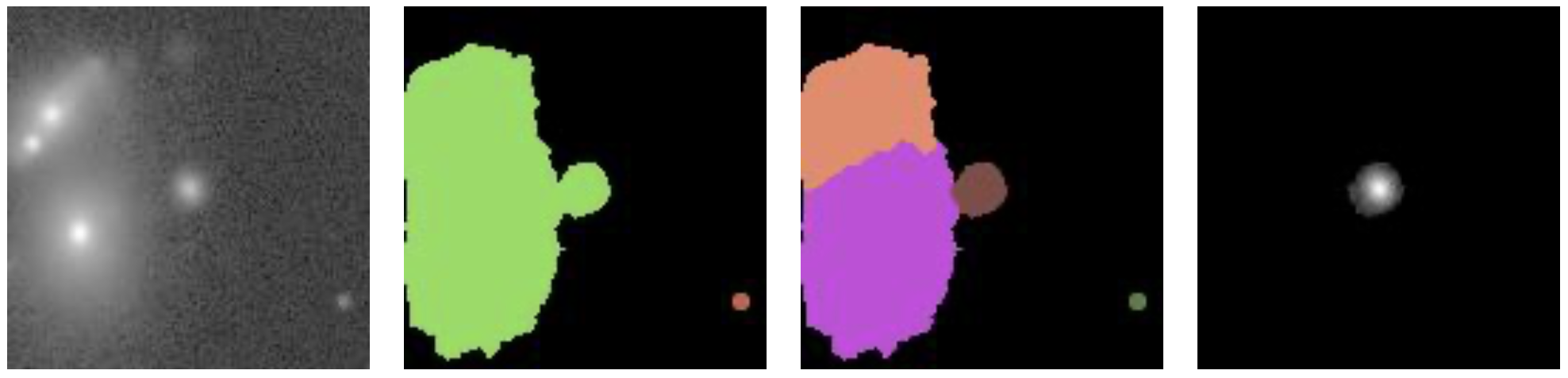}
    \caption{Example of the complete mask generation process. From left to right: the original mock image, the segmentation map before deblending, the segmentation map after deblending, and the image after applying the mask to isolate the central galaxy.}
    \label{fig:4}
\end{figure}

We detect galaxy light sources using the \pyth{detect_sources} function from \pyth{photutils.segmentation}, applying thresholds generated by \pyth{detect_threshold}. A minimum of eight connected pixels above the threshold is required for detection. Since many mock images contain blended sources, further separation is necessary to isolate the central galaxies for CAS parameter calculations. However, \pyth{detect_sources} identifies only regions above the threshold without deblending overlapping sources. To address this, we apply the \pyth{deblend_sources} function, which segments overlapping sources into individual components. After deblending, we obtain a full segmentation map in which each detected source and background is uniquely labelled.  This segmentation map provides the flexibility to construct different masks according to the requirements of subsequent analyses, for example, a mask containing only the central galaxy or a mask that combines the central galaxy with the background sky while excluding other sources. Figure~\ref{fig:4} illustrates the effect of each step in the mask–generation process, and the fourth panel presents an example of the final isolated central galaxy, produced by applying a mask that selects only the central galaxy region from the segmentation map.

\subsubsection{Asymmetry} \label{sec:2.4.2}
We calculate the asymmetry parameter, $A$, following \citet{conselice_2008_the}:
\begin{equation}
    A = \min \left( \frac{\sum{|I_0 - I_{180}|}}{\sum{|I_0|}} \right) - \min \left( \frac{\sum{|B_0 - B_{180}|}}{\sum{|I_0|}} \right)
\label{eqn:2.4}
\end{equation}
Here, $I_0$ represents the intensity of the original image, and $I_{180}$ is that of the image rotated by 180 degrees. The use of intensity means that this is a flux-weighted quantity. The rotational centre is defined as the point that minimises the asymmetry value, requiring asymmetry calculations across various centres to identify the minimum. We adopt a numerical optimisation approach using the \pyth{scipy.optimize.minimize} function to identify the rotational centre. The second term corrects for background noise, where $B_0$ and $B_{180}$ are the intensities from a blank sky region and its 180-degree rotation, respectively. This correction term is normalised by the original image's intensity and minimised using the same procedure as for the galaxy image. 

\begin{figure}
    \centering
    \includegraphics[width=0.47\textwidth]{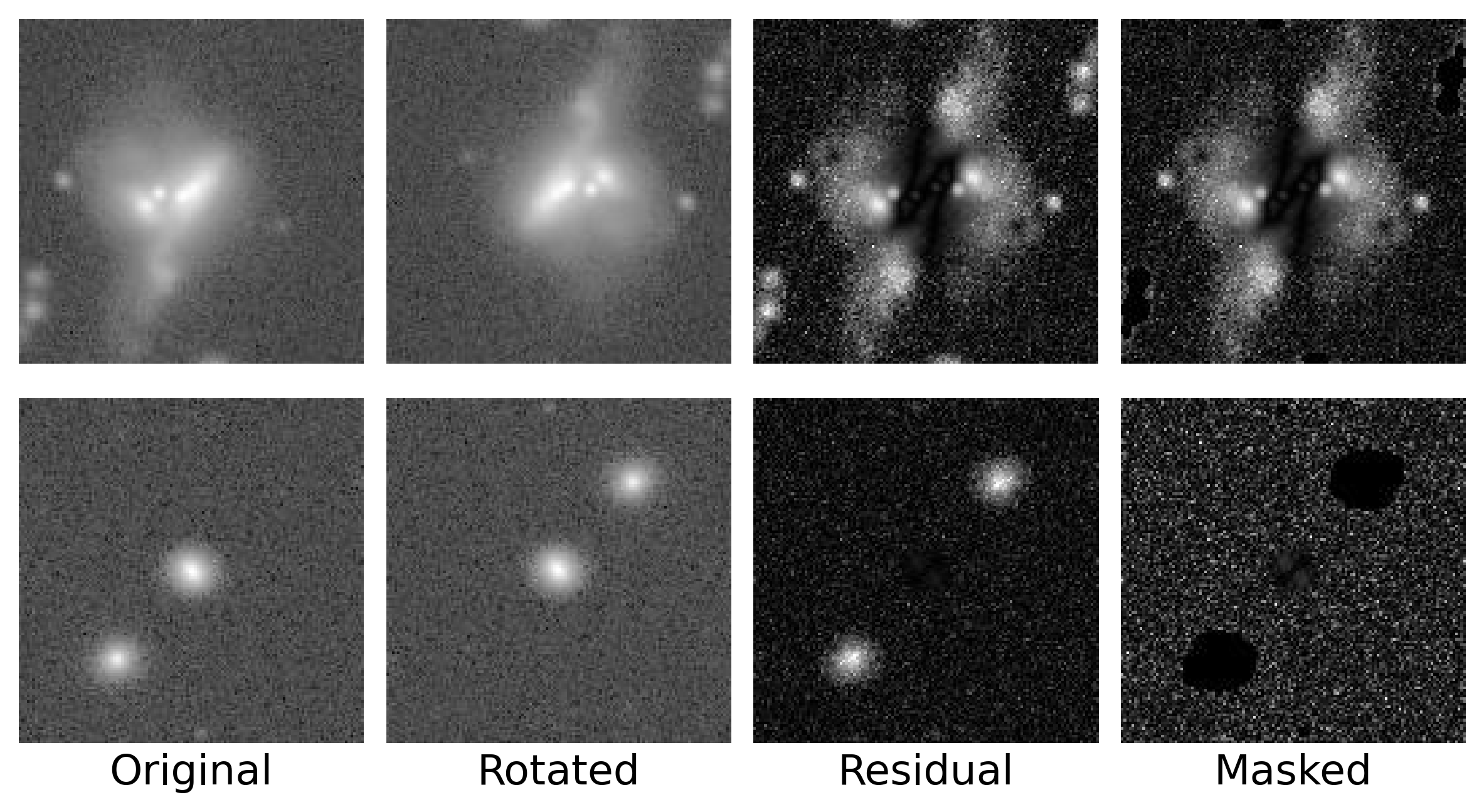}
    \caption{The asymmetry calculation process for a highly asymmetric galaxy (top row) and a more symmetric galaxy (bottom row). From left to right, the panels show: (1) the original image, (2) the $180^\circ$ rotated image, (3) the residual image calculated as the absolute difference between the original image and the rotated image, and (4) the masked residual image, where a binary mask combining the original and rotated masks is applied to exclude regions not associated with the central galaxy in either image. The asymmetry parameter is then calculated from the sum of the masked residual image. The top row corresponds to a highly asymmetric galaxy with $A = 0.69 \pm 0.0004$, while the bottom row shows a more symmetric galaxy with $A = 0.02 \pm 0.002$.
    }
    \label{fig:5}
\end{figure}

The residual image is computed as the absolute difference between the original image \(I_{0}\) and the 180-degree rotated image \(I_{180}\), as expressed in Equation~\ref{eqn:2.4}.  
This residual quantifies the asymmetry of the light distribution.  
To ensure that the measurement includes the central galaxy together with the surrounding sky background while excluding all other deblended sources, we construct masks including the central galaxy and the background sky region. Specifically, we start from the deblended segmentation map and assign a value of 1 to pixels belonging either to the central galaxy or to the empty sky, while all other sources remain 0. This mask is then rotated by \(180^{\circ}\) and logically intersected with the original mask so that only pixels present in both orientations are used when summing the residual flux. This procedure guarantees that the asymmetry is measured over the central galaxy and its adjacent background, but never contaminated by neighbouring objects. Figure~\ref{fig:5} illustrates this process, showing examples of a highly asymmetric galaxy and a more symmetric system.

For the background correction (the second term in Equation \ref{eqn:2.4}), \citet{conselice_2008_the} estimate the background using a blank sky region of the same size as the galaxy. However, as our mock images lack separate blank sky regions, we use the outskirts of the images instead. The background is identified using the 2D mask before deblending, where pixels labelled as 0 represent background regions free of light sources. We randomly sample the background pixels and fill them into the central galaxy region to create a matched background area. Its asymmetry is calculated using the same method as the uncorrected galaxy and subtracted to obtain the final background-corrected asymmetry. To reduce uncertainties from random sampling, we repeat the correction 100 times, taking the mean as the final value and the standard deviation as the uncertainty, which is also the main uncertainty source of the asymmetry parameters. Although more robust asymmetry metrics such as RMS Asymmetry \citep{2024OJAp....7E..77S} have been proposed to reduce noise sensitivity, our dataset consists of simulated mock images with uniform resolution and minimal noise contamination. Thus, we adopt the traditional asymmetry calculation method from \citet{conselice_2008_the}.

\subsubsection{Smoothness} \label{sec:2.4.3}
The calculation of the smoothness parameter, S, is based on Equation \ref{eqn:2.5} \citep{conselice_2008_the}. The core idea involves subtracting the smoothed images from the original images.
\begin{equation}
    S = 10 \left\{ \left( \frac{\sum (I_{x,y} - I_{x,y}^{\sigma})}{\sum I_{x,y}} \right) - \left( \frac{\sum (B_{x,y} - B_{x,y}^{\sigma})}{\sum I_{x,y}} \right) \right\}
\label{eqn:2.5}
\end{equation}
The light intensity at position $(x, y)$ in the original image is denoted as $I_{x,y}$, while $I_{x,y}^{\sigma}$ represents the intensity of the blurred image using a smoothing kernel of size $\sigma$. The smoothing kernel $\sigma$ is determined by the Petrosian radius $r_p$ of the galaxy \citep{1976ApJ...209L...1P}, given by $\sigma = 0.2 \times 1.5\, r_p(\eta = 0.2)$, where $r_p(\eta = 0.2)$ is the radius at which the ratio ($\eta$) of surface brightness at $r_p$ to the mean surface brightness within $r_p$ is 0.2 (e.g. \citealt{lotz_2004_a}). In this work, we compute $r_p$ directly from the 2D light distribution by measuring cumulative and local surface brightness in circular apertures centred on the galaxy, and identifying the radius where $\eta = 0.2$. Similar to the Asymmetry calculation, $S$ also accounts for background noise by applying the same smoothing kernel to blur the background intensity $B_{x,y}$ to $B_{x,y}^{\sigma}$ \citep{conselice_2003_the}.

\begin{figure}
    \centering
    \includegraphics[width=0.48\textwidth]{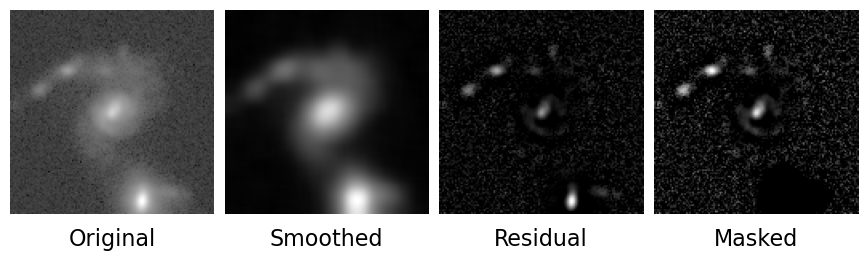}
    \caption{Example of generating a residual image for calculating the smoothness parameter, $S$. The first image shows the masked galaxy, the second image presents the smoothed version with a kernel size of 16 pixels ($r_p(\eta = 0.2) = 44.8$ pixels), and the third image shows the residual obtained by subtracting the smoothed image from the masked one. This galaxy has a smoothness value of $S = 0.34 \pm 0.0003$.}
    \label{fig:6}
\end{figure}

To smooth the mock image, we apply \pyth{scipy.ndimage.uniform_filter} with a kernel $\sigma = 0.3 r_p(\eta = 0.2)$. Residual images are then generated by subtracting the smoothed images from the original ones. As in the asymmetry calculation, to exclude contamination from other sources, we apply a mask that retains only the central galaxy and the background (from the deblended segmentation map). We also exclude the innermost circular region of radius \(0.25\,r_p(\eta=0.2)\), in order to avoid spurious residuals from the highly concentrated nuclear light \citep{lotz_2004_a}. The uncorrected smoothness is computed by dividing the sum of the residual pixel values by the sum of the original image's pixel values within the mask. Figure~\ref{fig:6} illustrates this sequence (for clarity, the central \(0.25\,r_p\) excision is not shown in the panels).

\section{Results} \label{sec:3}
This section presents the results obtained from the visual classification (Section \ref{sec:3.1}) and the two automated classification methods (Sections \ref{sec:3.2} and \ref{sec:3.3}). Because visual classification remains the most reliable approach for identifying low surface brightness tidal debris and broader interaction-driven disturbance signatures, we use it as the benchmark for evaluating the automated methods. We also compare the results of the two automated methods (Section \ref{sec:3.4}) and explore the relation with stellar mass across all classification methods (Section \ref{sec:3.5}).

\subsection{Visual Classification} \label{sec:3.1}
\begin{table}
\centering
\caption{Comparison of confidence-level distributions for the visual classification of interaction-driven disturbance signatures in the original sample from \citet{khalid_2023_characterising} and the remade mock-image sample used in this work.}
\begin{tabular}{lcccc}
\hline
Disturbance Category & \textbf{Conf. Level} & \textbf{Original (\%)} & \textbf{Remade (\%)} \\
\hline
\textbf{Total} & $0$ & $53.4 \pm 1.2$ & $64.8 \pm 1.1$  \\
& $1$ & $12.4 \pm 0.8$ & $10.1 \pm 0.7$ \\
& $2$ & $12.1 \pm 0.8$ & $7.3 \pm 0.6$ \\
& $3$ & $22.1 \pm 1.0$ & $17.8 \pm 0.9$\\
\hline
\textbf{Stream/Tail} & $0$ & $86.5 \pm 0.8$ & $95.7 \pm 0.5$\\
& $1$ & $4.9 \pm 0.5$ & $0.9 \pm 0.2$\\
& $2$ & $4.8 \pm 0.5$ & $1.2 \pm 0.3$\\
& $3$ & $3.7 \pm 0.4$ & $2.2 \pm 0.3$\\
\hline
\textbf{Shell} & $0$ & $99.2 \pm 0.2$ & $99.8 \pm 0.1$ \\
& $1$ & $0.4 \pm 0.2$  & $0.1 \pm 0.1$\\
& $2$ & $0.1 \pm 0.1$ & $0.0 \pm 0.0$\\
& $3$ & $0.3 \pm 0.1$ & $0.2 \pm 0.1$\\
\hline
\textbf{Asymmetric Halo} & $0$ & $60.4 \pm 1.1$ & $67.9 \pm 1.1$\\
& $1$ & $11.7 \pm 0.8$ & $9.7 \pm 0.7$\\
& $2$ & $10.7 \pm 0.7$ & $7.2 \pm 0.6$ \\
& $3$ & $17.3 \pm 0.9$ & $15.3 \pm 0.8$\\
\hline
\textbf{Double Nucleus} & $0$ & $81.2 \pm 0.9$ & $89.4 \pm 0.7$\\
& $1$ & $3.3 \pm 0.4$ & $1.8 \pm 0.3$ \\
& $2$ & $4.9 \pm 0.5$ & $1.6 \pm 0.3$\\
& $3$ & $10.7 \pm 0.7$ & $7.3 \pm 0.6$\\
\hline
\end{tabular}
\label{table:2}
\end{table}

Of the 1,826 mock galaxy images, 17.8\% (324 galaxies) were assigned a final confidence level of 3, 7.3\% (134 galaxies) a final confidence level of 2, 10.1\% (185 galaxies) a final confidence level of 1, and 64.8\% (1,183 galaxies) a final confidence level of 0. We measured the distribution of galaxies across these confidence levels for each visually classified disturbance category. If galaxies with a confidence level of 2 or higher are considered likely to host a visually identifiable interaction-driven disturbance, then 25.1\% of the galaxies in the remade mock-image sample fall into this category. Table \ref{table:2} summarises the confidence-level distributions for the individual disturbance categories, together with the corresponding classifications from \citet{khalid_2023_characterising}. Uncertainties for each confidence level are estimated using the beta distribution, which provides reliable estimates even for small and imbalanced datasets \citep{Cameron_2011}.

Compared to the original classification results, the overall frequency of visually identified disturbance signatures decreases in the remade mock-image sample. These reductions are primarily attributed to the increased distance and lower resolution of the remade images, which reduce the visibility of faint structures and make morphological classification more ambiguous. In particular, cosmological surface-brightness dimming makes low-surface-brightness structures substantially harder to detect. Stream/tail features are most strongly affected, because their narrow and extended morphologies are especially sensitive to losses in resolution and surface brightness \citep{martin_2022_preparing}. Among the four visually classified categories, asymmetric haloes remain the most commonly identified disturbance signature, followed by double nuclei. Stream/tail features are relatively rare, while shell features are nearly absent. This is consistent with the classification framework described in Section~\ref{sec:2.2}: asymmetric haloes and double nuclei trace broader disturbance signatures, whereas streams/tails and shells are specific low surface brightness tidal debris structures that are more strongly affected by surface-brightness dimming and resolution loss. This trend is qualitatively consistent with the findings of \citet{khalid_2023_characterising}.

\subsection{SSL Model} \label{sec:3.2}
\citet{desmons_2023_detecting}'s SSL model consists of two elements: a self-supervised encoder used for pre-training and a linear classifier used for binary classification. In this work, the binary labels correspond to the presence or absence of interaction-driven disturbances, as defined in Section~\ref{sec:2.3}.

\begin{figure}
    \centering
    \includegraphics[width=0.48\textwidth]{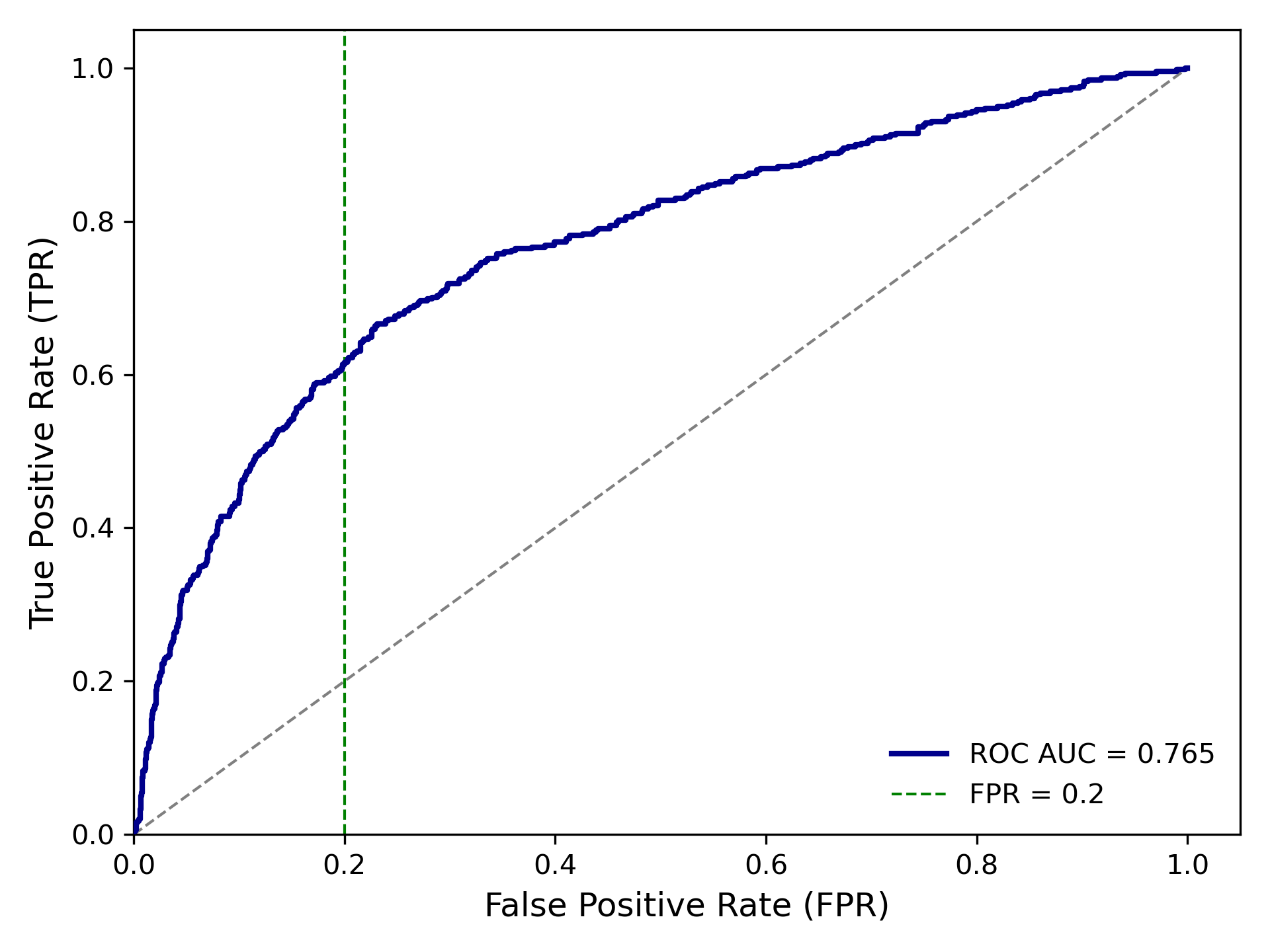}
    \caption{The ROC curve shows the performance of the originally trained model on the entire mock image dataset. Each point on the curve represents the True Positive Rate (TPR) and False Positive Rate (FPR) calculated at a specific threshold. Varying the threshold divides the continuous linear classifier scores into binary labels, with each point on the curve reflecting the corresponding FPR and TPR values derived from these binary labels. }
    \label{fig:7}
\end{figure}

We first apply \citet{desmons_2023_detecting}'s originally trained model directly to our mock image dataset. This results in a less-than-ideal performance, with a ROC AUC of 0.765 (Figure \ref{fig:7}) and completeness (TPR) of only 61\%, while maintaining a low contamination rate of 20\%. 

\begin{figure}
    \centering
    \includegraphics[width=0.48\textwidth]{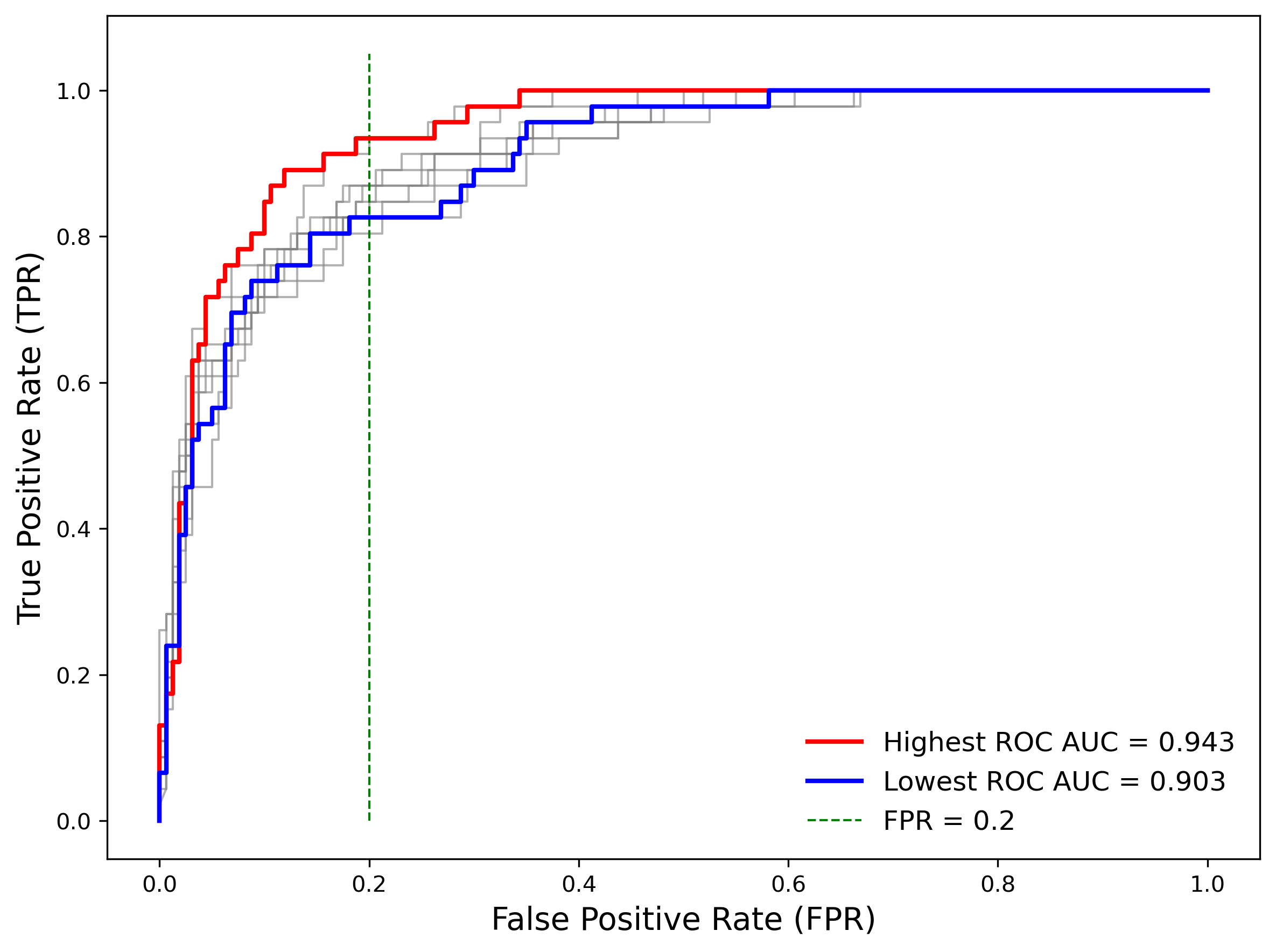}
    \caption{The ROC curve for ten re-trained linear classifiers, trained on 1440 mock images and tested on 206 galaxies. The model with the highest ROC AUC is shown in red, the lowest is shown in blue, and the other nine models are shown in grey.}
    \label{fig:8}
\end{figure}

Since the self-supervised model was originally trained using a very large unlabelled dataset of about 44,000 galaxies, we do not have enough mock images to retrain the self-supervised model. We explore using the original self-supervised model and retraining the linear classifier on the mock image dataset to determine how well the model can be adapted to new data. The originally trained self-supervised model is used as the encoder when retraining the linear classifier with our mock images, followed by validation and testing to evaluate performance. The labelled mock image dataset, consisting of 1,826 images, was annotated using visual classification. The dataset is divided into training, validation, and test sets with ratios of 78.8\%, 9.9\%, and 11.3\%, respectively. This results in 1,440 images for training, 180 for validation, and 206 for testing. We fine-tune hyperparameters by adjusting training epochs, testing different learning rates, modifying augmentation techniques, and applying early stopping to optimise the re-trained model. The linear classifier with the best hyperparameters is re-trained ten times on the training dataset and tested on the test dataset. Then we use the ROC curve to evaluate the performance of the re-trained models on the new mock image dataset. Figure \ref{fig:8} shows the ROC curve for the ten re-trained linear classifiers. 

We highlight the model that has the highest ROC AUC in red, with ROC AUC=0.943. The final performance of the ten models was evaluated by calculating the mean and standard deviation of their results. This gives an average final ROC AUC = $0.915 \pm 0.013$. We choose the same threshold as \citet{desmons_2023_detecting} that contamination (FPR) is 0.2, which leads to an average TPR = $0.859 \pm 0.039$. The best model has TPR = 0.935 when FPR = 0.2. In the case of only retraining the linear classifier on the new dataset, the SSL model still achieved good completeness (TPR) for low contamination (FPR).These values refer to the experiments using the full training set of 1,440 images. 

\begin{figure}
    \centering
    \includegraphics[width=0.48\textwidth]{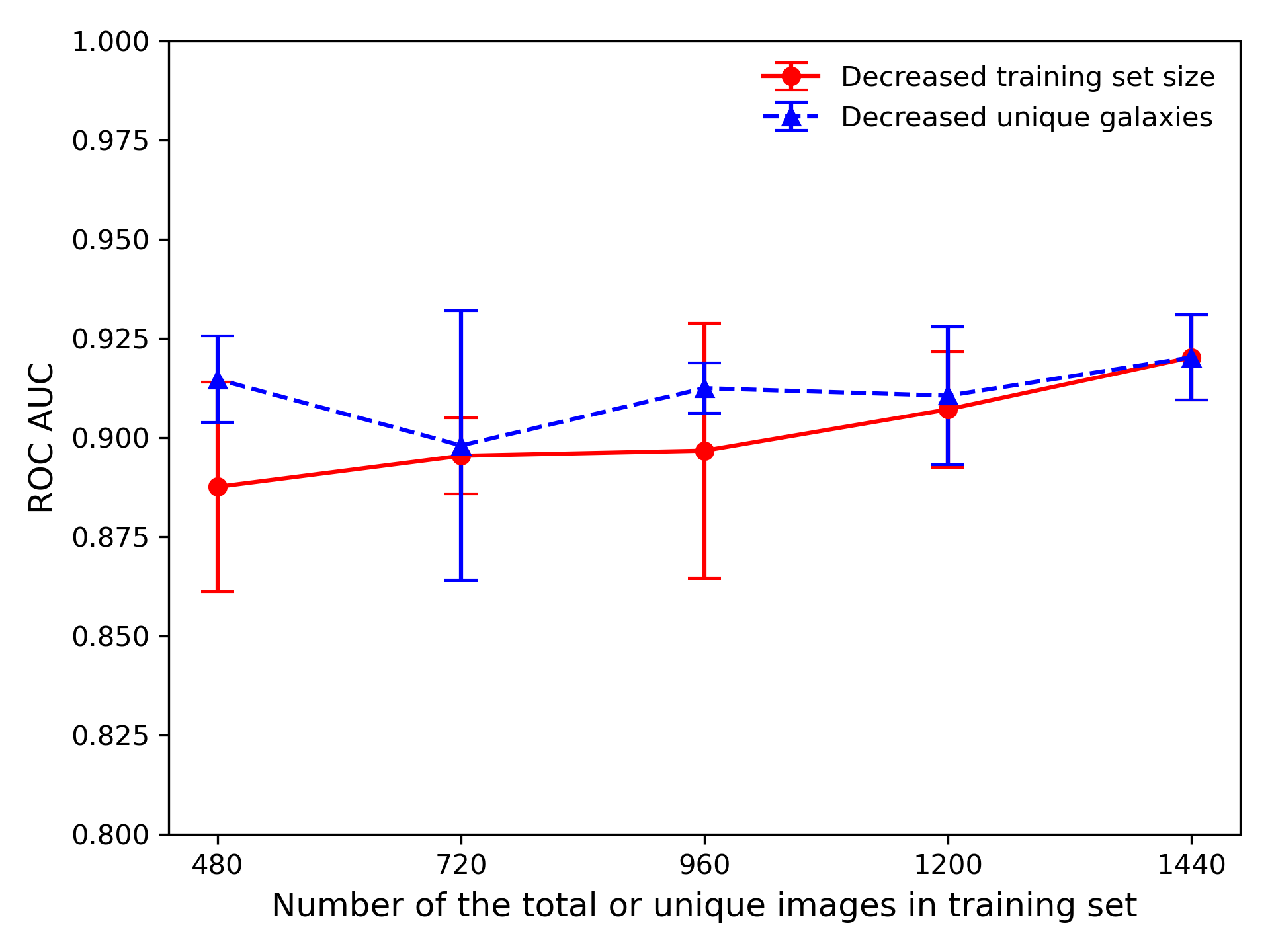}
    \caption{ROC AUC on the test set when reducing the training set size compared to reducing only the number of unique galaxies while maintaining the overall training set size. For each training set size, we trained 10 models and took the mean value as the final result, and the standard deviation around this was used as the uncertainty.}
    \label{fig:9}
\end{figure}

One of the reasons the SSL model is so attractive is its ability to achieve high completeness with a small labelled sample. We therefore explore here how reducing the training dataset affects the re-trained classifier’s ability to adapt to the new dataset. To do this, we reduced the training dataset size from 1,440 to 480 mock images, with a step size of 240, while keeping the validation and test sets unchanged. Additionally, following \citet{desmons_2023_detecting}, we tested reducing the number of unique galaxies in the training set while repeating some or all of these galaxies to maintain the original training set size (1,440 images). Figure \ref{fig:9} compares the ROC AUC scores for reducing the training set size versus reducing the number of unique galaxies in the training set.

As shown, reducing the size of the training set does lead to a decrease in the model's performance, although the decrease was not substantial. When the training set size was reduced from 1,440 to one-third (480), the ROC AUC falls from $0.915 \pm 0.013$ to $0.888 \pm 0.026$. The re-trained model maintains higher performance when we decrease the number of unique galaxies from 1440 to 480, the mean ROC AUC is almost the same value. 

We find that keeping the size of the training set unchanged but changing the unique images in it gives slightly better results than simply reducing the training set. However, this difference is not significant, only $0.025$ in average ROC AUC. With ROC AUC = $0.913 \pm 0.011$, the training set with only 480 unique mock images has almost the same result as the training set with 1440 unique images. The SSL model only needs a relatively small number of new images to be re-trained for a new dataset and achieve good performance.

\begin{figure}
    \centering
    \includegraphics[width=0.48\textwidth]{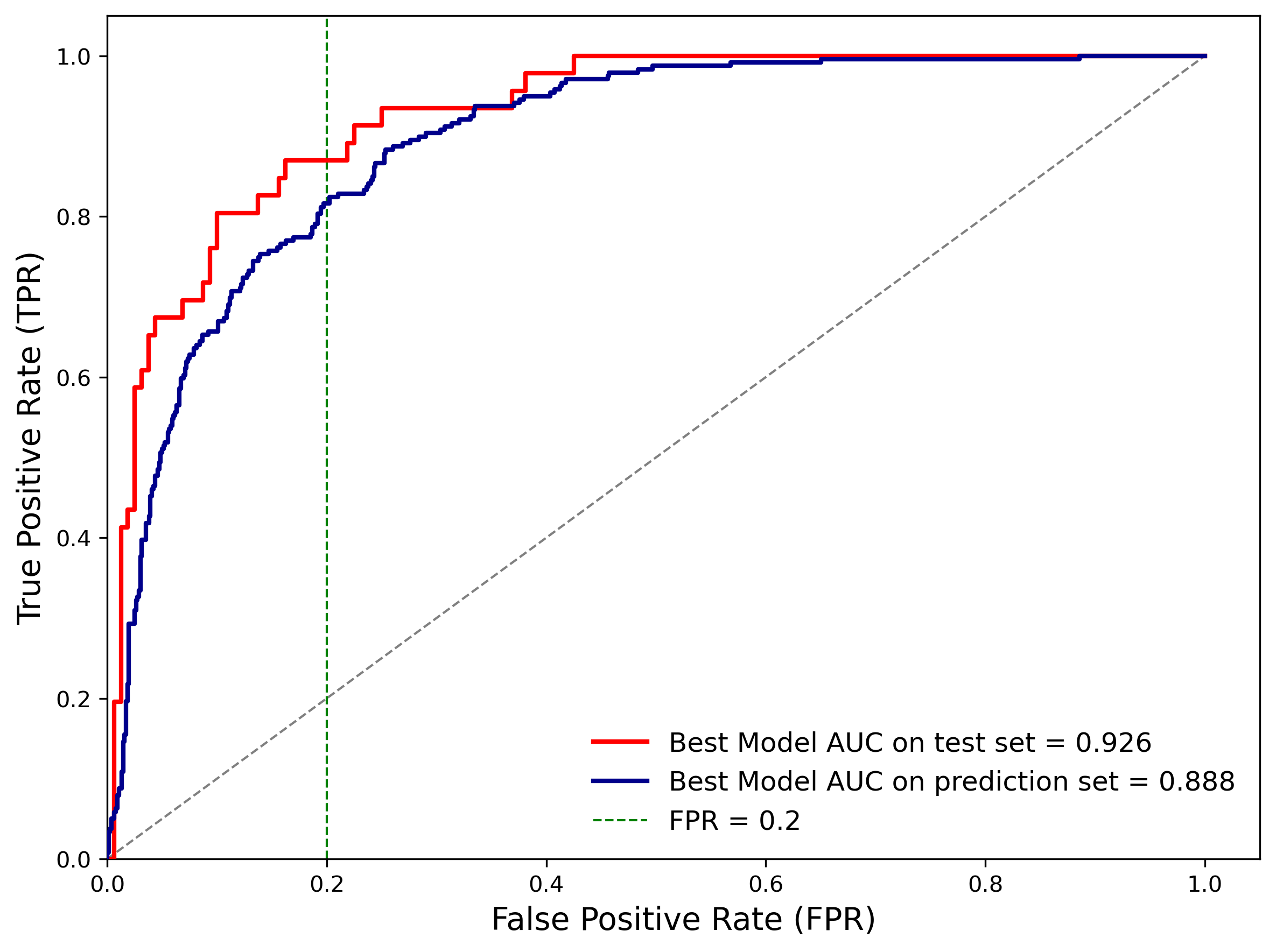}
    \caption{ROC curves for the best-performing linear classifier evaluated on two datasets: the test set of 206 images and the prediction set of 960 previously unseen images. The model achieves a ROC AUC of 0.926 on the test set, which decreases to 0.888 when applied to the prediction set. }
    \label{fig:10}
\end{figure}

In the previous evaluation, we split the total of 1,826 mock images into training, validation, and test sets and calculated the average ROC AUC and recall on the test set. To better compare with the CAS results across more galaxies, it is more appropriate to apply the model to a dataset that was not used during training, validation, or testing. This allows us to compute evaluation metrics for this unseen dataset and generate a classifier score for each individual image. Therefore, we selected the best-performing re-trained linear classifier (highest ROC AUC = 0.926) from the experiments using a training set that included only 480 unique galaxies (with 180 images for validation and 206 for testing). This model was then applied to the prediction set of the remaining 960 mock images that were not used in the training or validation process. Figure \ref{fig:10} shows the ROC curve change from the test set to the prediction set. When applied to the prediction set, both the ROC AUC and recall (TPR) at FPR = 0.2 decreased. The ROC AUC decreased from 0.926 to 0.888, and the recall at FPR = 0.2 dropped from 0.87 to 0.80. Since the test sets are much smaller than the prediction dataset, the distribution of data is different between the test sets and the prediction set, which is likely the reason behind these reductions.  Using the FPR = 0.2 threshold, we also calculated a precision of 0.56 and an F1-score of 0.65. These metrics, along with the linear classifier scores from this model, are compared to the CAS method results on this prediction set in the next Section. 

\begin{figure}
    \centering
    \includegraphics[width=0.42\textwidth]{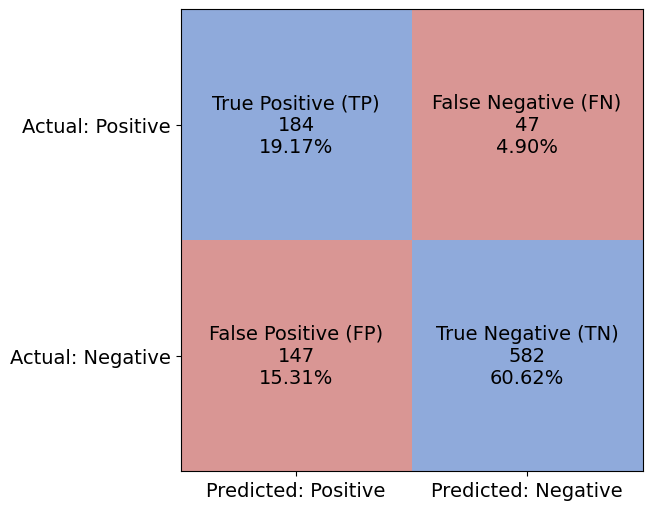}
    \caption{Confusion matrix of the re-trained SSL model evaluated on the prediction dataset with the threshold set to the classifier score corresponding to FPR = 0.2 on the ROC curve. True Positive and True Negative cases, representing correctly classified samples, are shown in blue, while False Negative and False Positive cases, representing misclassified samples, are shown in red. The metrics of recall, precision, and F1-score are derived from these four classification outcomes.}
    \label{fig:11}
\end{figure}

Figure \ref{fig:11} presents the confusion matrix for the re-trained SSL model tested on the prediction dataset. We use true positives (TP), true negatives (TN), false positives (FP), and false negatives (FN) in the confusion matrix to compute key evaluation metrics, including recall, precision, and the F1-score. The model correctly classifies 184 galaxies as disturbed, consistent with the visual classifications, resulting in a high recall. However, it also misclassifies 147 galaxies as disturbed, leading to a lower precision.

\subsection{CAS parameters} \label{sec:3.3}
\begin{figure}
    \centering
    \includegraphics[width=0.47\textwidth]{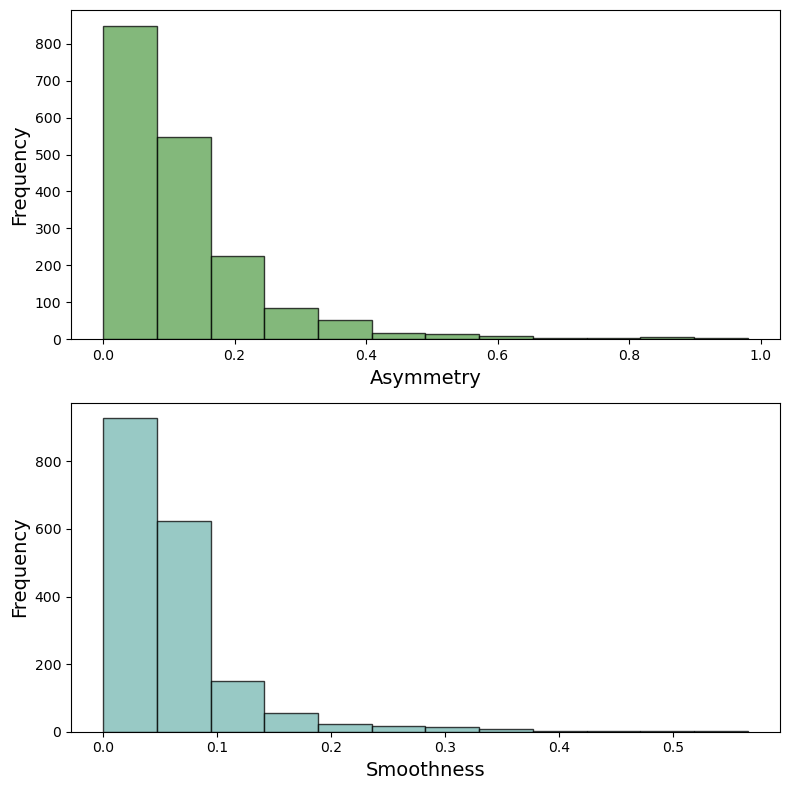}
    \caption{The distribution of the asymmetry and smoothness parameters for the full dataset.}
    \label{fig:12}
\end{figure}

As described in Section \ref{sec:2.4}, we calculate the asymmetry and smoothness parameters for the full dataset in order to apply a CAS-based selection of disturbed or merging systems. Figure \ref{fig:12} illustrates the distribution of the asymmetry and smoothness parameters for the mock images.

\begin{figure}
    \centering
    \includegraphics[width=0.48\textwidth]{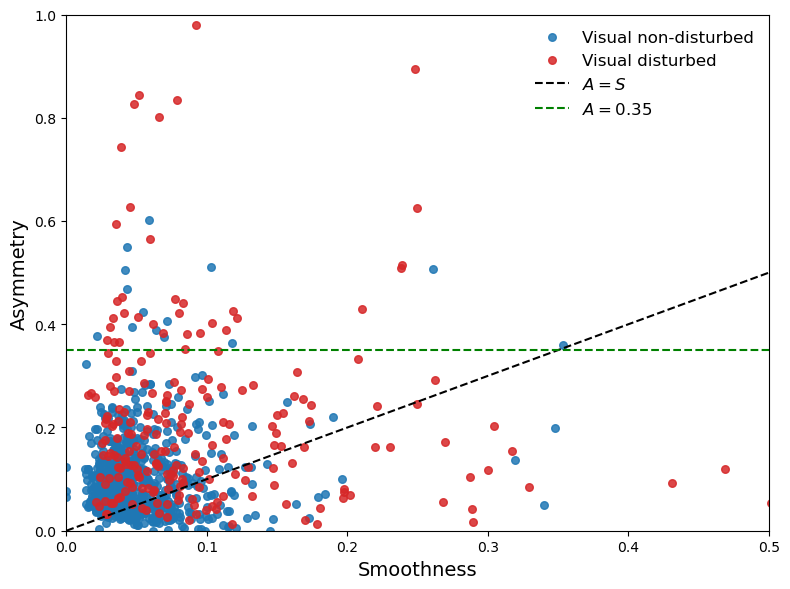}
    \caption{The Smoothness-Asymmetry space with galaxies coloured by visual classification: red points represent visually classified disturbed systems, and blue points represent visually classified non-disturbed systems. The green dashed line indicates the $A = 0.35$ threshold commonly used in CAS-based merger selection, and the black dashed line shows the $A = S$ boundary.}
    \label{fig:13}
\end{figure}

We first applied the galaxy-merger criteria from \citet{conselice_2003_the}, which define mergers as systems with $A > 0.35$ and $A > S$. Based on these criteria, we classified the galaxies into CAS-selected mergers and non-mergers. Figure \ref{fig:13} presents the smoothness-asymmetry (S-A) space, with each point coloured according to its visual classification: red for visually classified disturbed systems and blue for visually classified non-disturbed systems. The green dashed line denotes the threshold at $A = 0.35$, while the black dashed line shows the $A = S$ boundary. As $A > S$ does not provide a strong discriminant for this sample, we focus on $A > 0.35$ as the CAS-based merger criterion in subsequent analysis.

\begin{figure}
    \centering
    \includegraphics[width=0.42\textwidth]{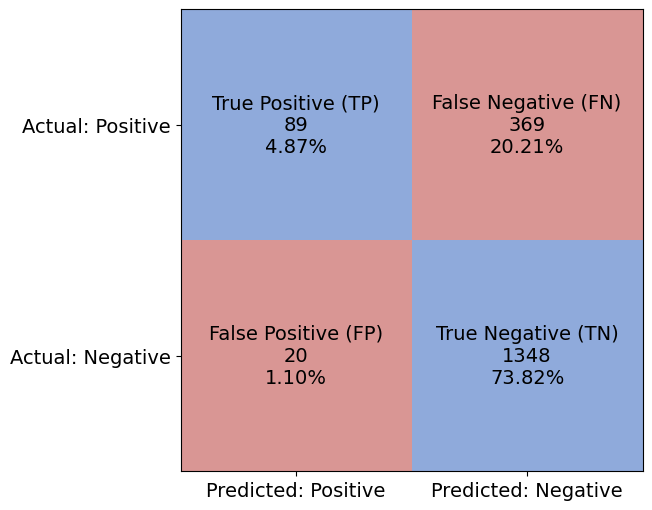}
    \caption{Confusion matrix of the CAS parameter results. True Positive and True Negative cases, representing correctly classified samples, are shown in blue, while False Negative and False Positive cases, representing misclassified samples, are shown in red. The metrics of recall, precision, and F1-score are derived from these four classification outcomes.}
    \label{fig:14}
\end{figure}

Of the 1,826 galaxies analysed, the CAS parameters identified 89 galaxies as mergers, representing $5.97 \pm 0.5\%$ of the total sample. To evaluate the classification performance of the CAS method consistently, we applied the same metrics used for the self-supervised learning models, taking the visual classification results as the benchmark. The resulting precision was 0.82, the recall was 0.19, and the F1 score was 0.31. Figure~\ref{fig:14} presents the confusion matrix for the CAS results.

The high precision indicates that most galaxies identified by the CAS parameters as mergers are indeed visually classified as disturbed. However, the low recall of 0.19 indicates that the CAS method fails to recover a substantial fraction of the galaxies visually classified as disturbed. Specifically, of the 458 galaxies visually classified as disturbed, only 89 were correctly identified by the CAS parameters. Conversely, among the 109 galaxies predicted as mergers by CAS, 89 were true positives. This highlights the method's strength in minimising false positives, but also its limitation in completeness. Figure \ref{fig:13} illustrates this issue clearly: many visually classified disturbed systems lie below the $A = 0.35$ threshold and are therefore not recovered by the standard CAS criterion.

To facilitate a direct comparison with the SSL model, we evaluated the performance of the CAS method on the same prediction set of 960 images described in Section~\ref{sec:3.2}. The CAS method demonstrated consistently stable performance, with a precision of 0.77, a recall of 0.20, and an F1-score of 0.32 even on this smaller sample. 

\subsection{Re-trained SSL Model Vs. CAS parameters} \label{sec:3.4}
Table \ref{table:3} shows the metrics obtained from the best-performed re-trained classifier and the CAS parameters on the prediction dataset. 

\begin{table}
\centering
\caption{Performance comparison between the Re-trained SSL model and CAS parameters on the prediction set.}
\centering
\begin{tabular}{cccc}
\hline\hline
\textbf{Metric} & \textbf{Re-trained SSL} & \textbf{CAS ($A > 0.35$)} \\  \hline
Accuracy & 0.80 & 0.79 \\
Precision & 0.56 & 0.77 \\
Recall & 0.80 & 0.20 \\
F1 Score & 0.65 & 0.32 \\
\hline\hline
\end{tabular}
\label{table:3}
\end{table}

The re-trained SSL model and the CAS-based classifiers exhibit similar accuracy scores, with the SSL model achieving 0.80 and the CAS threshold of \(A > 0.35\) yielding 0.79. This indicates that the two methods have similar overall `correctness' when classifying galaxies as disturbed or non-disturbed under the adopted benchmark. However, it is important to note that our mock image dataset is imbalanced. According to the visual classifications, only 458 out of 1826 galaxies (\(\sim25\%\)) are classified as disturbed (confidence level \(\geq 2\)), and in the prediction set, only 231 out of 960 galaxies fall into this category. As a result, accuracy alone cannot fully reflect the method’s performance, since predicting most instances as negative would still yield a high score. To provide a more meaningful assessment, we therefore focus on precision, recall, and F1-score in the following comparisons.

In examining precision and recall, we observe a clear trade-off between the re-trained SSL model and the CAS method under the threshold of \(A > 0.35\). The CAS parameters have higher precision, achieving 0.77 compared to the re-trained SSL model's 0.56, suggesting that CAS is more conservative and produces a cleaner sample of disturbed systems with fewer false positives. However, this comes at the cost of low recall, where the re-trained SSL model shows a significant advantage with a recall of 0.80, capturing a majority of the galaxies with interaction-driven disturbances, whereas the CAS method reaches only 0.20 and therefore misses a large fraction of them. Overall, the CAS method favours purity over completeness, while the re-trained SSL model favours completeness and recovers a substantially broader disturbed population, albeit with more false positives.

\begin{figure}
    \centering
    \includegraphics[width=0.47\textwidth]{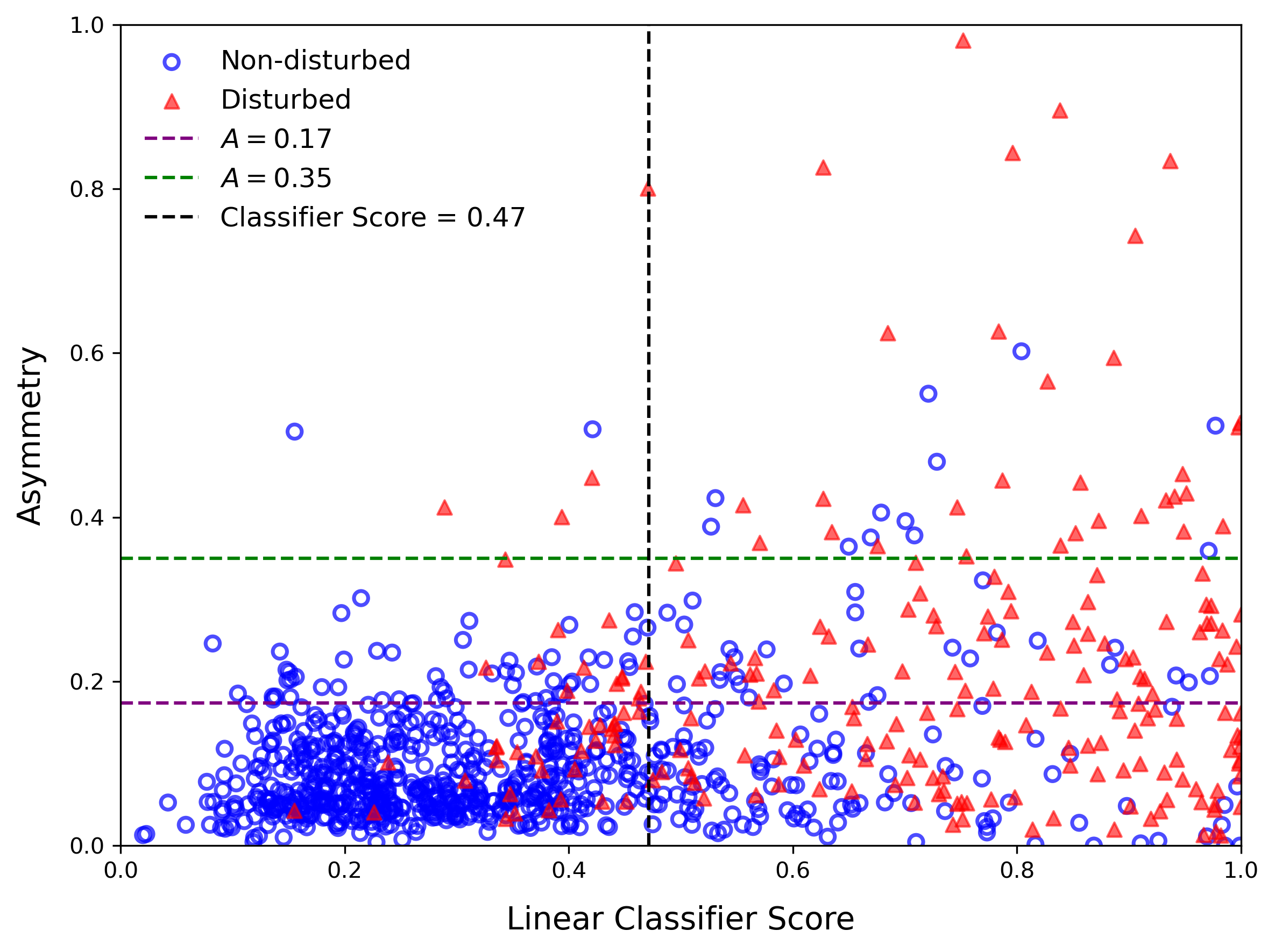}
    \caption{Asymmetry parameter as a function of linear classifier scores from the best model. Classifier Score = 0.47 is the threshold for this best model at FPR = 0.2. $A > 0.35$ is the typical value used to classify galaxy mergers in the CAS method, while $A > 0.14$ is the asymmetry threshold driven from the ROC Curve. The labels are based on the visual disturbance classifications.}
    \label{fig:15}
\end{figure}

Figure \ref{fig:15} shows the model's linear classifier score versus the asymmetry parameter, with visually classified disturbed systems shown by the red triangles. This clearly shows the different focus of these two methods. Most of the galaxies with $A > 0.35$ are visually classified as disturbed. However, there are also many interaction-driven disturbances below the \(A = 0.35\) line, which means they are missed by the traditional CAS method, resulting in high precision but notably low recall. In our best re-trained SSL model, the linear classifier score of 0.47 is the threshold to predict disturbed systems in the prediction set: galaxies with classifier scores above this value are classified as having interaction-driven disturbances. In Figure \ref{fig:15}, a large fraction of the points labelled as disturbed lie in the region with classifier score $\geq 0.47$. However, this region also contains a considerable number of visually non-disturbed galaxies. This distribution corresponds to the high recall and lower precision shown in Table \ref{table:3}. Note that the threshold of the classifier score = 0.47 is only for this linear classifier, the threshold for every re-trained model would be different, even if they are trained with the same hyperparameters and datasets. Also, the CAS threshold depends on the resolution and depth of the images being analysed, and a given threshold should not be taken as an absolute for a different set of images.

\begin{figure}
    \centering
    \includegraphics[width=0.47\textwidth]{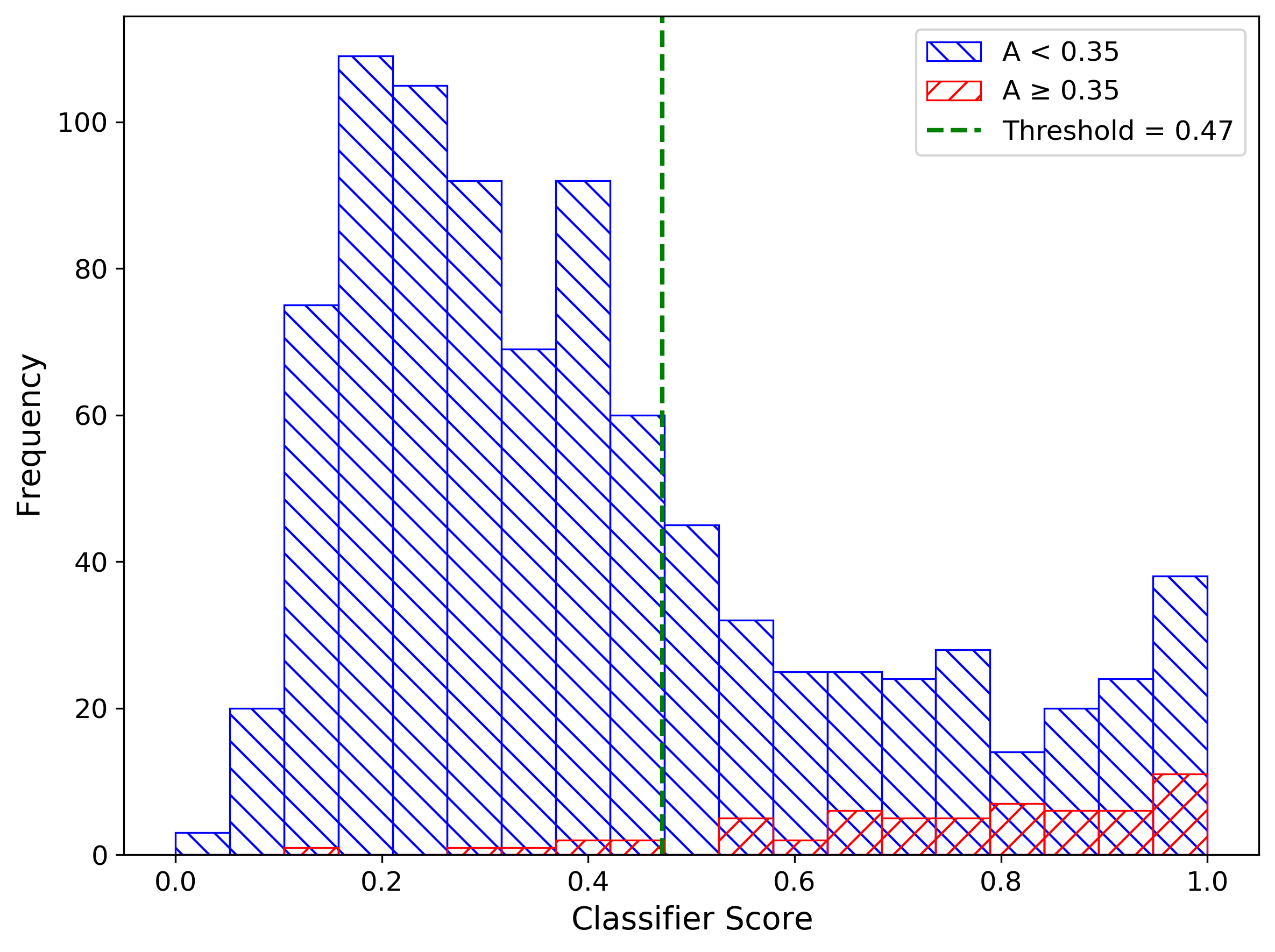}
    \caption{The distribution of classifier scores with classification results from CAS parameters. Galaxies with $\text{Asymmetry} > 0.35$ are classified as merging galaxies by the CAS parameters.}
    \label{fig:16}
\end{figure}

Figure \ref{fig:16} shows the distribution of classifier scores for the different classes identified by the CAS parameters with a threshold of \(A > 0.35\). Most systems selected as mergers by the CAS method also have classifier scores above the SSL threshold, reinforcing that the re-trained SSL model performs better in capturing more galaxies with disturbances.

The F1 score indicates that the re-trained SSL model is a more balanced classification method, with an F1 score of 0.68 compared to 0.28 for the CAS method. This difference suggests that the re-trained SSL model is more effective in comprehensively identifying disturbances when both precision and recall are considered together. The re-trained SSL model detects as many true positives as possible while maintaining a reasonable false positive rate. In contrast, the CAS parameters are more conservative, favouring precision over recall. While this reduces false positives, it limits the overall sensitivity of the CAS method, resulting in lower effectiveness when evaluated across both precision and recall.

The conservative behaviour of the CAS method is likely to be a result of the intrinsic design of the classical asymmetry parameter. Because the asymmetry parameter is a flux-weighted method, the bright central regions of galaxies dominate the measurement, while faint extended structures contribute only minimally. As a result, diffuse low surface brightness tidal features tend to produce only weak asymmetry residuals, leading to systematically low recall even when threshold values are varied (see Section \ref{sec:4.3}).

Alternative asymmetry measures specifically developed for detecting faint tidal debris include shape asymmetry \citep{10.1093/mnras/stv2878}, which computes the asymmetry on binary detection masks and therefore gives equal weight to all morphological components, and isophotal asymmetry methods \citep{sazonova2025statmorphlsstquantifyingcorrectingmorphological}, which evaluate asymmetry across multiple surface-brightness contours, are more sensitive to diffuse tidal features.

This explains why the CAS method yields high precision but low completeness, while the re-trained SSL classifier recovers a much larger fraction of visually identified disturbance hosts.

\subsection{Stellar Mass Relation} \label{sec:3.5}
\begin{figure}
    \centering
    \includegraphics[width=0.47\textwidth]{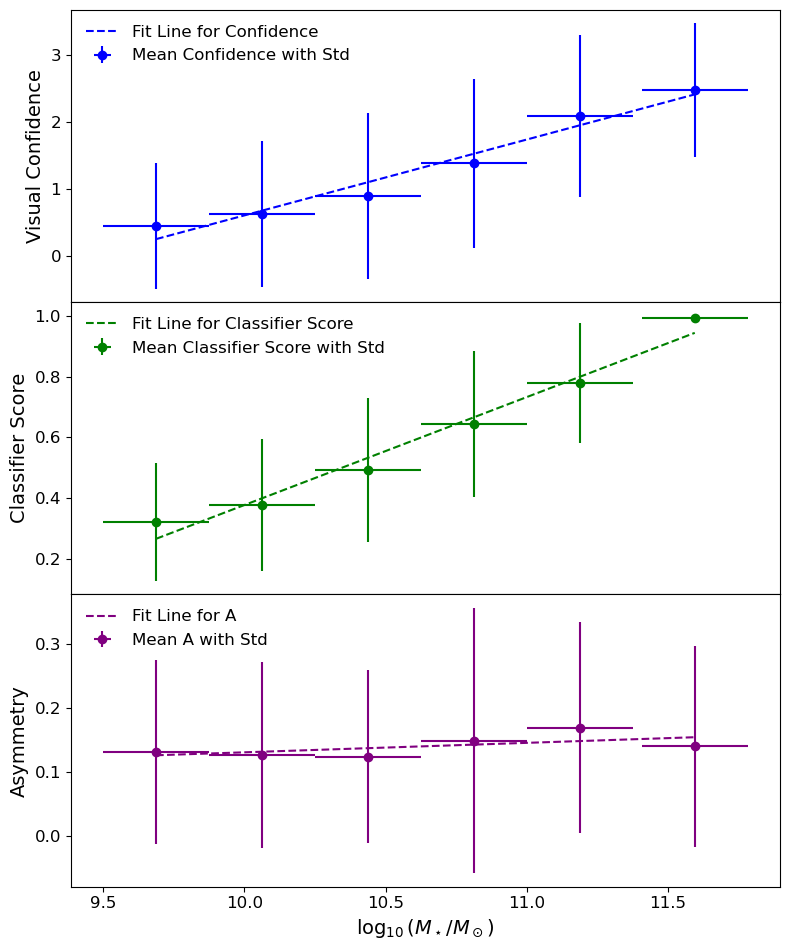}
    \caption{The results obtained from visual classification (top panel), our re-trained SSL model (middle panel) and the Asymmetry parameter (bottom panel) as a function of stellar mass. Stellar mass is binned in intervals of 0.375 dex in $\log_{10}(M/M_\odot)$; each point represents the mean value within a bin. Horizontal error bars denote the bin width (0.375 dex), while vertical error bars represent the standard deviation within each bin. }
    \label{fig:17}
\end{figure}

It has been shown in both observations and simulations that the incidence of tidal debris and other merger-related disturbance signatures depends on stellar mass (e.g. \citealt{atkinson_2013_faint}, \citealt{blek_2020_census}, \citealt{martin_2022_preparing}, \citealt{desmons_2023_detecting}, \citealt{khalid_2023_characterising}). To examine the impact of the classification approach on the relationships found, we analysed the stellar mass dependence of results from the three classification methods: visual classifications, the re-trained SSL model, and CAS parameters. Figure \ref{fig:17} shows the visual classification confidence levels, classifier scores from the re-trained SSL model and the asymmetry parameter as functions of $\log_{10} \left( M_\star/M_\odot \right)$. We analyse the relationship between the SSL classifier score and stellar mass using the model's performance on the unseen prediction set.

All three methods show generally increasing trends with stellar mass. Table~\ref{table:4} shows the best-fit lines for all three methods, along with their corresponding Spearman $p$-values. Both the visual classification and the SSL classifier show strong positive correlations between higher stellar mass and a higher likelihood of identifying disturbed systems, as shown by their sharp best-fit lines and extremely small Spearman \textit{p}-values (\(p < 10^{-45}\)), indicating statistically significant correlations. In contrast, the asymmetry parameter displays a nearly flat best-fit slope and a small but formally significant \textit{p}-value of (\(6.31 \times10^{-5}\)). Although this formally meets the threshold for statistical significance, the effect size is negligible. This highlights the conservative behaviour of the classical asymmetry criterion and suggests that the CAS method is much less responsive to the stellar-mass trend seen in the broader visually disturbed population than the visual and SSL-based approaches.

\begin{table}
\caption{Best-fit linear relations between each classification metric and the logarithm of stellar mass, along with the corresponding p-values from the Spearman rank correlation test. The statistically significant correlations are highlighted in bold font.}
\centering
\begin{tabular}{ccc}
\hline\hline
\textbf{Metric} & \textbf{Best fit line} & \textbf{Spearman \textit{p}-value}\\ \hline
Visual        & $y = (1.13 \pm 0.11) x - (10.70 \pm 1.19)$  & $1.51 \times 10^{-46}$ \\ 
SSL       & $y = (0.36 \pm 0.03) x - (3.18 \pm 0.31)$  &$7.89 \times 10^{-61}$              \\ 
CAS          & $y = (0.015 \pm 0.01) x - (0.017 \pm 0.10)$  &$6.31 \times10^{-5}$                \\ 
\hline\hline
\end{tabular}
\label{table:4}
\end{table}

\section{Discussion} \label{sec:4}
Here we summarise the main findings of this analysis, including a comparison with results from previous research. 

\subsection{Visual Classifications} \label{sec:4.1}
In Section \ref{sec:3.1}, we compared the visual classification results of the original mock images from \citet{khalid_2023_characterising} with the visual classifications of the remade images in this paper. Both sets represent the same galaxies but differ in the distance from the observer and the image resolution. We found that these factors have a significant impact on the visual classifications, reducing the overall fraction of visually identified disturbance signatures in the remade images relative to \citet{khalid_2023_characterising}. However, the distributions between the different disturbance morphologies are qualitatively retained.

We also compare our visual classification results with those from previous observational studies examining similar samples (i.e. size, distance from the observer and survey depth). \citet{atkinson_2013_faint} visually classified 1781 galaxies from CFHTLS with a redshift range of $0.02 < z < 0.4$. Our mock image sample closely matches in size and spans a similar redshift range. They classified these galaxies into five confidence levels, ranging from 0 to 4, which differs slightly from our confidence levels of 0 to 3. However, combining their confidence levels 3 and 4, we find strong agreement between the visual classification results across all of their confidence levels as seen in Table \ref{table:5}.

\begin{table}
\caption{Comparison of the confidence-level distributions between \citet{atkinson_2013_faint} and this analysis.}
\centering
\begin{tabular}{ccc}
\hline\hline
\textbf{Confidence} & \textbf{\citet{atkinson_2013_faint}} & \textbf{This paper} \\ \hline
$\geq$ 3       & $17.6 \pm 1.0$         & $17.8 \pm 0.9$ \\ 
= 2       & $7.6 \pm 0.7$                 & $7.3 \pm 0.6$                 \\ 
= 1         & $11.3 \pm 0.8$                 & $10.1 \pm 0.7$                 \\ 
= 0       & $64.4 \pm 1.9$                & $64.8 \pm 1.1$                    \\ 
\hline\hline
\end{tabular}
\label{table:5}
\end{table}

\citet{Huang_2022} conducted a visual classification of 2,649 images from the Wide layer of the HSC-SSP third data release. Their tidal feature fraction for prominent tidal features is $0.28 \pm 0.01$. Since their study focused on prominent tidal features, we primarily compare their results to our confidence level 3 fraction ($0.18 \pm 0.01$). This is significantly lower than their reported tidal features. Even considering galaxies with a confidence level $\geq 2$ in our study, which has a fraction of $0.25 \pm 0.02$, our results are still slightly lower. This difference is likely due to the disparity in stellar mass: their sample focuses on galaxies with stellar masses $M_{\star} > 10^{11} M_{\odot}$, while our stellar masses range from $3.2 \times 10^{9} M_{\odot}$ to $6.5 \times 10^{11} M_{\odot}$. As discussed in Section \ref{sec:4.4}, the incidence of identified disturbance signatures shows a positive correlation with stellar mass, with higher-mass galaxies exhibiting a higher disturbance fraction.
 
\citet{desmons_2023_galaxy} visually classified a sample of 852 galaxies from the Galaxy And Mass Assembly survey (GAMA; \citealt{driver_2011_galaxy}) using a classification scheme closely aligned with the one adopted here. They found an overall tidal feature fraction of $f_{\text{tidal}} = 0.23 \pm 0.02$. Because their classification framework grouped together both specific tidal debris and broader disturbance signatures such as asymmetric haloes and double nuclei, this quantity is most directly comparable to our confidence level $\geq 2$ fraction ($0.25 \pm 0.02$), which in the framework of this paper represents the fraction of galaxies likely to host a visually identifiable interaction-driven disturbance.

For comparison with other observational and simulation works, it is important to consider the detectability of tidal features within our sample. \citet{khalid_2023_characterising} used a toy model to estimate the detectability of tidal features in mock images similar to those used in this study, as a function of the stellar mass of the tidal feature and the amount of contaminating mass from nearby galaxies. They found that \textsc{TNG100} LSST-like mock images at $z = 0$ are able to resolve tidal features with stellar masses of approximately $10^8\ M_\odot$, even when dispersed over regions as large as 2500 kpc$^2$, corresponding to a surface brightness of 29.7 mag arcsec$^{-2}$, provided that contaminating light from surrounding stellar mass does not exceed 29.4 mag arcsec$^{-2}$. 

However, since the galaxies in our mock sample are at $z \sim 0.2$, surface-brightness dimming must also be taken into account. At this redshift, the dimming effect is approximately 0.79 mag arcsec$^{-2}$, reducing the maximum area over which a tidal feature can be spread and still be visible in our images to $\sim$2300 kpc$^2$, assuming no more than 30.1 mag arcsec$^{-2}$ of contaminating surface brightness. These constraints imply that many faint or diffuse tidal features—especially in lower-mass galaxies—may fall below the detection threshold, helping to explain the lower visually identified disturbance fractions observed in the remade mock images, as quantified in Table \ref{table:2}.

Despite these limitations, our classifications remain consistent with previous observational studies. In general, the comparison with previous visual-classification studies shows strong consistency, demonstrating the reliability of our visual-classification criteria and methods. This suggests that our visual classification is effective in identifying interaction-driven disturbance signatures in mock images and serves as a robust benchmark for evaluating other automated detection methods.

We note that our classification criterion for double nuclei is intentionally conservative: requiring additional evidence of faint connecting or surrounding disturbance reduces contamination from chance projections, but may exclude genuine merging systems in which a double nucleus is not accompanied by visible low-surface-brightness tidal debris. This is particularly relevant for early-type or gas-poor systems, where interactions between elliptical galaxies can produce a double nucleus without generating strong morphological disturbances in the outer halo. If the visual labels produced here are used as ground truth for future deep learning training, this should be taken into account, as a model trained on these labels may learn to require the co-presence of other disturbance features to identify double-nucleus systems, potentially leading to an underestimate of genuine double-nucleus mergers lacking such signatures.

\subsection{Machine Learning Results} \label{sec:4.2}
We compare the results from our linear classifier re-trained on the mock images with those from \citet{desmons_2023_detecting}. In our model, the encoder remains the original from \citet{desmons_2023_detecting}. Thus, we aim to explore how the model’s performance changes when using a new dataset, with only the classifier component re-trained.

We achieved a high ROC AUC of $0.92 \pm 0.01$ on the test set after retraining on 1,440 mock images, which aligns well with the results from \citet{desmons_2023_detecting} (average ROC AUC = $0.91 \pm 0.002$). However, their linear classifier was trained on only 600 images. When our classifier was trained on 720 images, it achieved an average ROC AUC of $0.90 \pm 0.03$, still consistent with \citet{desmons_2023_detecting}'s model. Although the model re-trained on 1,440 mock images achieved similar overall ROC performance, there was a noticeable difference in TPR at a fixed FPR = 0.2. The original model showed a higher TPR of $0.94 \pm 0.1$, while our re-trained model reached $0.85 \pm 0.04$. Training on 720 images further decreased the TPR to $0.82 \pm 0.03$. This suggests that, although the re-trained model maintains overall ROC performance, it is less effective in identifying true positives at lower FPR thresholds.

The performance of the SSL model on a new dataset further illustrates the value of self-supervised learning relative to fully supervised machine-learning approaches for identifying disturbed systems. \citet{wechsler_2018_the} used a Convolutional Neural Network (CNN) to identify tidal features in 1,781 galaxies that were visually classified by \citet{atkinson_2013_faint}. Using a similar-sized dataset to ours, they achieved a completeness of 0.76 at a contamination of 0.20, while our SSL model has a higher completeness of $0.85 \pm 0.04$. \citet{hdomnguezsnchez_2023_identification} used the same CNN structure as \citet{wechsler_2018_the} but trained on a much bigger dataset with 6,000 mock images. They have a similar recall as our SSL model, with a value of 0.85, but a slightly higher precision of 0.72 compared to our 0.64. Given that their training set is substantially larger than ours, this further highlights the advantage of the SSL approach: even with a relatively small labelled training set, it retains a strong ability to identify a large fraction of the disturbed population.

\citet{hdomnguezsnchez_2023_identification} tested their model, trained on mock images, on a new dataset of real HSC-SSP images. However, the model's performance decreased significantly, achieving only an ROC AUC of 0.64. This trend aligns with the performance drop observed when we applied \citet{desmons_2023_detecting}'s originally trained model to our mock image dataset without retraining the linear classifier, where the ROC AUC decreased to 0.765. Similar performance declines were also noted by \citet{pearson_2019_identifying} when models were directly applied to different datasets.

In general, the original \citet{desmons_2023_detecting}'s model, which uses the same dataset for both the self-supervised model and the linear classifier, does show a better performance. However, we achieved good performance using the original self-supervised model and retraining the linear classifier, with only a slight performance gap compared to the original model. This provides valuable insights for future classification tasks using the SSL model, including applying it to the new LSST survey. Instead of retraining the entire SSL model on a new dataset, one can simply retrain the linear classifier using a smaller labelled dataset to achieve strong performance. The original encoder can still complete its task well on the new dataset, which can save the time needed to use a larger unlabelled dataset to train the self-supervised model.

\subsection{Choice of Asymmetry Thresholds} \label{sec:4.3}
\begin{figure}
    \centering
    \includegraphics[width=0.47\textwidth]{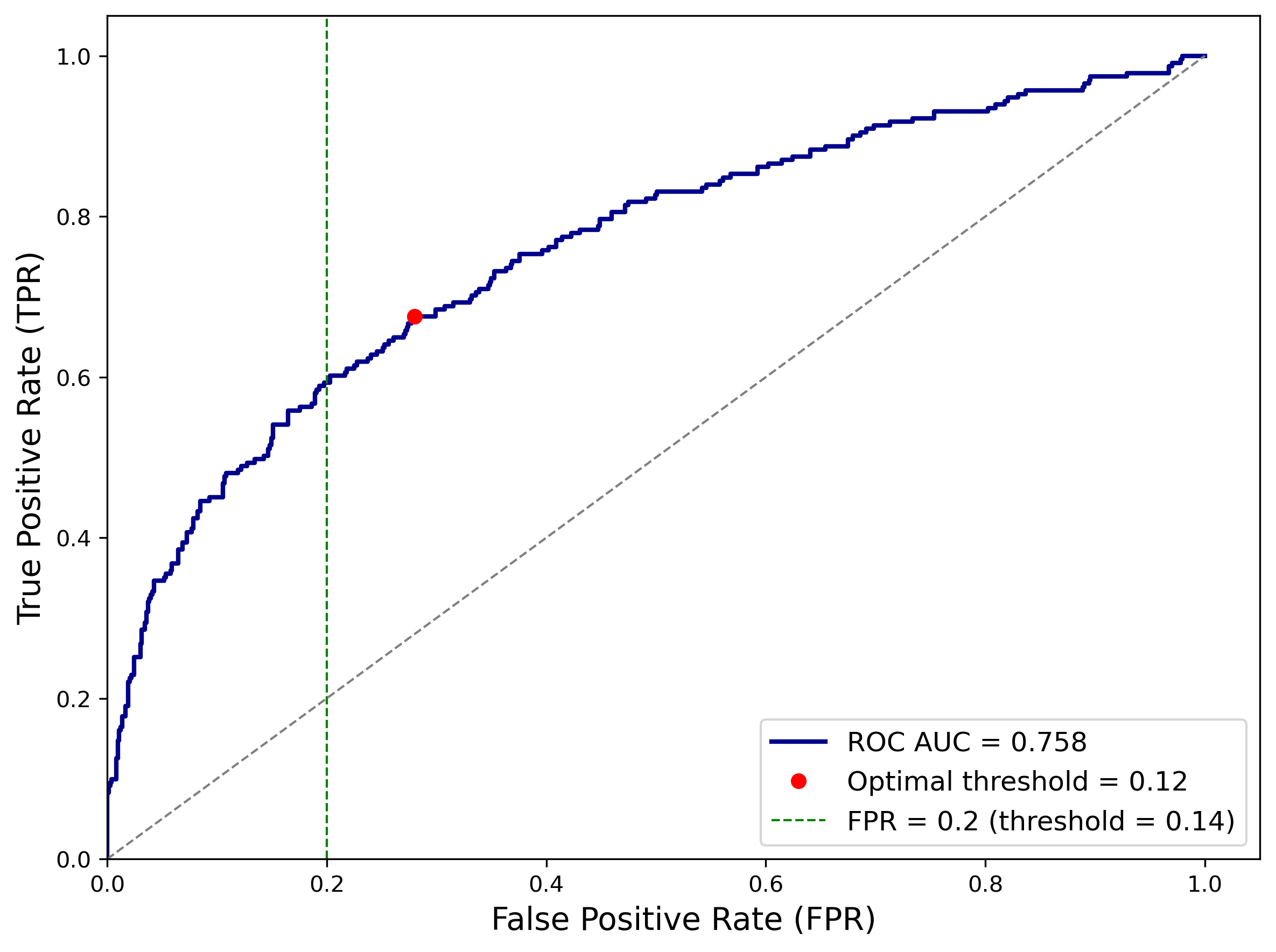}
    \caption{ROC curve for the asymmetry parameter on the prediction set of 960 galaxies with ROC AUC = 0.758. The red point minimizes the distance to the ideal classifier \((\mathrm{FPR} = 0,\, \mathrm{TPR} = 1)\), corresponding to an ROC-derived optimal threshold of \(A = 0.12\). The vertical dashed line indicates the point where \(\mathrm{FPR} = 0.2\), which corresponds to a threshold of \(A = 0.14\).}
    \label{fig:18}
\end{figure}

While the comparison between the CAS method and the SSL model in Section \ref{sec:3.4} adopts the traditional threshold of \(A > 0.35\), alternative thresholds may offer different trade-offs between precision and recall.  Since the SSL model is evaluated with an emphasis on completeness (recall), and the fixed CAS threshold of \(A > 0.35\) yields high precision but low recall, we refine the asymmetry threshold using the ROC curve. Figure~\ref{fig:18} shows the ROC curve for the CAS parameter method on the prediction set of 960 galaxies, which has an ROC AUC of 0.758.

In the ROC curve framework, a perfect classifier is represented by the point \((\mathrm{FPR} = 0,\, \mathrm{TPR} = 1)\). By identifying the point on the ROC curve that is closest to this ideal coordinate, we can determine the optimal classification threshold. Applying this approach to the prediction set of 960 galaxies, we identified an optimal asymmetry threshold of \(A > 0.12\), as shown in Figure~\ref{fig:18}. At this threshold, the CAS parameters method achieves an accuracy of 0.70, a precision of 0.42, a recall of 0.68, and an F1-score of 0.52. 

We also evaluated the CAS parameters method at the point where the false positive rate (FPR) equals 0.2, which corresponds to a threshold of \(A = 0.14\) to align with the evaluation criteria adopted for the SSL model. This alignment enables a consistent and fair comparison between the classification results of the CAS-based method and the SSL model. At \(A > 0.14\), the classifier achieves an accuracy of 0.75, precision of 0.48, recall of 0.59, and an F1-score of 0.53. Since the threshold is derived from the ROC curve, which plots the true positive rate (recall) against the false positive rate, the optimal threshold tends to prefer a higher recall, resulting in higher recall and lower precision compared to the threshold selected at FPR = 0.2.

Compared to the fixed criterion of \(A > 0.35\), the ROC-derived thresholds significantly improve recall at the cost of precision. Specifically, recall increases from 0.19 (for \(A > 0.35\)) to 0.68 (for the optimal ROC threshold \(A > 0.12\)) but is 0.59 for the FPR = 0.2 threshold \(A > 0.14\). Precision respectively decreases from 0.82 to 0.42 and 0.48. As a result, the F1-score improves from 0.31 to 0.52 and 0.53. This trade-off is consistent with the nature of the ROC curve, which focuses on maximising the true positive rate. As a result, the ROC-based threshold provides a more balanced sensitivity for identifying merger candidates, albeit at a cost of higher false positives.

\begin{figure}
    \centering
    \includegraphics[width=0.47\textwidth]{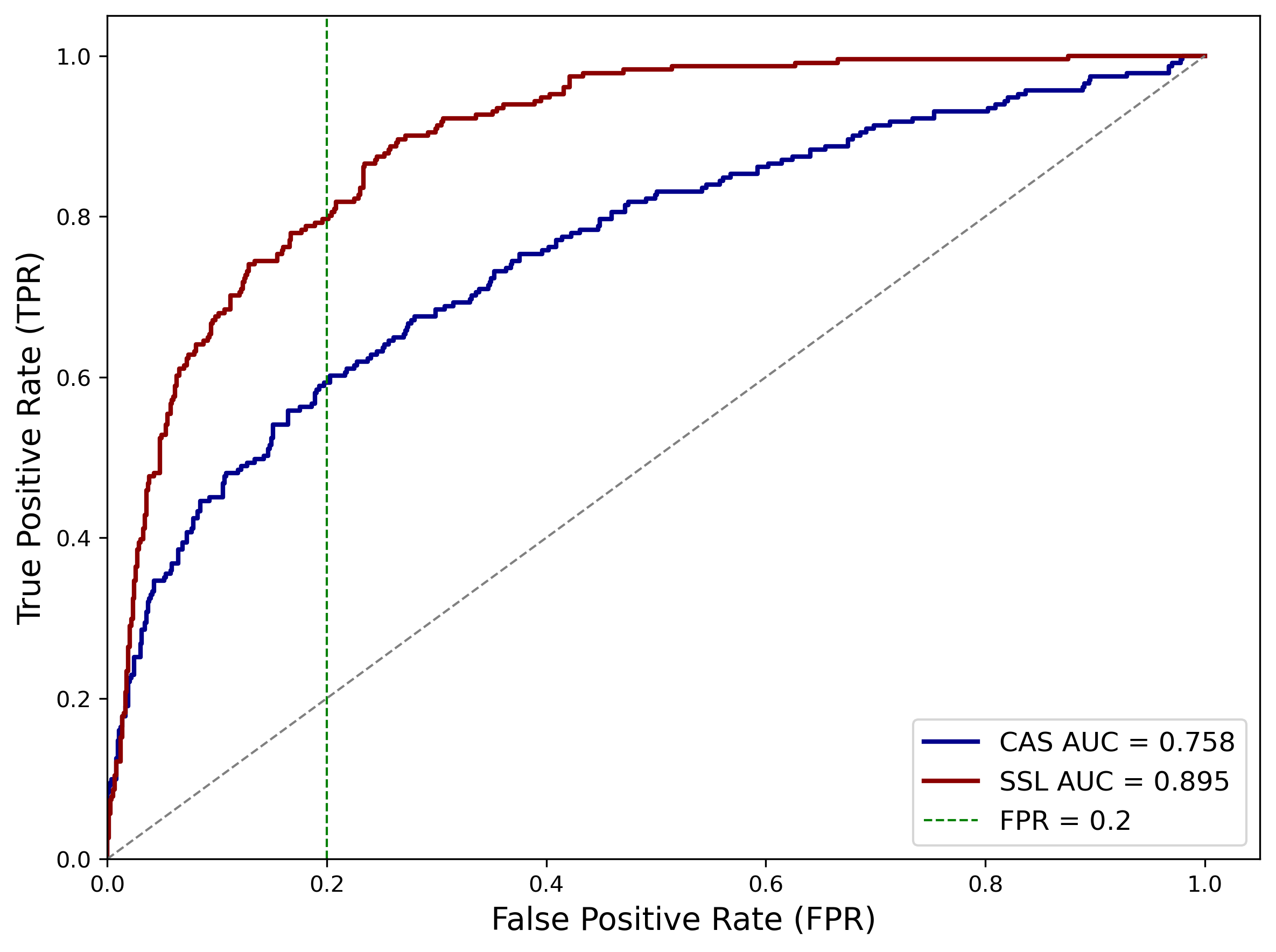}
    \caption{ROC curves for the re-trained SSL model (red) and the CAS parameter (blue) evaluated on the prediction set of 960 galaxies. The area under the curve (AUC) is 0.895 for the SSL model and 0.758 for the CAS method. The green dashed line marks the point where the false positive rate (FPR) equals 0.2, which is used as the classification threshold for comparison.}
    \label{fig:19}
\end{figure}

In the case of the SSL model, the classification threshold is selected at the point where the false positive rate (FPR) equals 0.2. Although both methods adopt the same thresholding strategy, the SSL model consistently outperforms the CAS approach across all evaluation metrics. Figure \ref{fig:19} presents a comparison of the ROC curves for the re-trained SSL model and the CAS parameters. The SSL model achieves a higher AUC of 0.90 compared to 0.76 for the CAS method, indicating a stronger overall classification performance. The CAS method, with a threshold \(A > 0.14\), achieves a recall of 0.59 and a precision of 0.48, resulting in an F1-score of 0.53. In contrast, the SSL model reaches a recall of 0.80 and a precision of 0.56, yielding a higher F1-score of 0.65. This demonstrates that the SSL classifier not only identifies more true tidal features but also maintains a lower false positive rate, leading to a more balanced and robust overall performance. The performance gap may arise from the limited descriptive power of the CAS parameters method, which relies solely on the asymmetry parameter and may not capture the full morphological complexity of tidal structures.  While the CAS threshold of \(A > 0.14\) offers a more balanced trade-off than the traditional \(A > 0.35\) threshold by improving recall, it still has a lower performance than the SSL model across all evaluation metrics, and \(A > 0.35\) instead represents a different trade-off compared to the SSL model, hence, \(A > 0.35\) was adopted in the previous comparison in Section \ref{sec:3.4}.

\subsection{Stellar Mass Relation} \label{sec:4.4}
In Section \ref{sec:3.5}, we find that both the visual classification and the re-trained SSL model exhibit strong positive correlations between stellar mass and the likelihood of identifying disturbed systems.This trend is consistent with previous simulations, which have generally found that more massive galaxies are more likely to host tidal features and other interaction-driven disturbance signatures. \citet{martin_2022_preparing} reported a similar correlation in the \textsc{NewHorizon} simulation, and \citet{khalid_2023_characterising} found a similar correlation in \textsc{NewHorizon}, \textsc{Eagle}, \textsc{IllustrisTNG}, and \textsc{Magneticum}.

Observationally, \citet{desmons_2023_detecting} visually classified galaxies in the HSC-SSP survey and found that the fraction of galaxies exhibiting tidal features increases with stellar mass. Similarly, \citet{kadofong_2018_tidal} found that tidal features are more prevalent in massive galaxies, reinforcing the connection between mergers and stellar mass. \citet{atkinson_2013_faint} further demonstrated that linear features, shells, and diffuse fans preferentially occur in galaxies with $M_{\star} \geq 10^{10.5} M_{\odot}$. Additionally, \citet{blek_2020_census} conducted a census of tidal features in the MATLAS (Mass Assembly of early-Type GaLAxies with their fine Structures; \citealt{10.1093/mnras/stu2019}) survey and confirmed a significant positive correlation between stellar mass and tidal fraction, with more massive galaxies exhibiting the highest rates of tidal features. \citet{Sola2022} characterised low surface brightness structures in annotated deep images of nearby massive galaxies at surface brightness limits of $\sim$28.3--29 mag arcsec$^{-2}$, comparable to those expected from early LSST operations, finding that more massive galaxies are twice as likely to host tidal debris. \citet{Sola2025} further confirmed, using deep CFHT imaging of 475 nearby massive galaxies across diverse environments, that stellar mass is the dominant factor driving tidal feature prevalence.

In contrast, the asymmetry parameter shows a much weaker relationship with stellar mass. As shown in Table~\ref{table:4}, the slope of the best-fit line is close to zero, and the corresponding \textit{p}-value of \(6\times10^{-5}\) is formally significant but reflects only a negligible effect. This weak trend suggests that the CAS method is less sensitive to stellar mass in our dataset, possibly due to its conservative nature and limited sensitivity to faint tidal features and other low-contrast disturbances, which may not significantly influence the flux-weighted asymmetry value. These results highlight that, while the visual classification and the SSL model both capture the increasing detectability of disturbance signatures in more massive galaxies, the asymmetry parameter provides much weaker discriminative power under the current imaging conditions.

We find a positive correlation between stellar mass and the likelihood of detecting interaction-driven disturbances at the surface-brightness depth of our mock images ($\mu_{\mathrm{lim}} \sim 30.3\ \mathrm{mag\ arcsec^{-2}}$), consistent with previous simulation and observational results. However, this trend likely reflects detection bias rather than intrinsic differences in merger activity. Tidal features in lower-mass galaxies tend to be fainter and more diffuse, making them harder to detect under fixed surface-brightness limits. As noted by \citet{martin_2022_preparing}, analysis of the \textsc{NewHorizon} cosmological simulations shows that tidal features become increasingly detectable across all galaxy masses as detection limits are lowered, suggesting that the observed stellar-mass dependence may arise because the detection limit prevents us from seeing fainter disturbance signatures in lower-mass galaxies. Additionally, the detectability of tidal structures likely depends not only on the host galaxy's stellar mass but also on the brightness and morphology of the features themselves. While we do not quantify feature brightness here, future work using asymmetry residuals or Sérsic model fits (e.g. \citealt{sazonova_2021_are}) may help disentangle these effects.

\section{Conclusions} \label{sec:5} 
We present three different methods for classifying and identifying interaction-driven disturbance signatures around galaxies and apply them to mock images from the \textsc{IllustrisTNG-100} simulation for comparison. By comparing the classification results from visual classification, an existing SSL model, and CAS parameters, we draw the following conclusions:

\begin{itemize}
\item The overall fraction of galaxies with a visually identified disturbance signature at confidence levels $\geq 2$ is $25.1 \pm 1.5 \%$, which is lower than the $34.2 \pm 1.8 \%$ fraction measured for the original mock images, which were placed closer to the observer. Galaxy distance impacts the completeness of visual classification. However, our visual classification results show good consistency with previous similar analyses.
\item By only retraining the linear classifier element of the \citet{desmons_2023_detecting}'s SSL model on a small labelled dataset (1440 images) of new mock images, the SSL model achieves a ROC AUC of $0.92 \pm 0.01$, showing good performance. This demonstrates excellent classification ability. Moreover, as the size of the dataset used to retrain the linear classifier is reduced, the model continues to demonstrate strong performance down to a sample of only 480 galaxies.
\item The re-trained SSL model performs comparably to the original model when using a similar-sized training set (720 images), achieving a ROC AUC of $0.90 \pm 0.03$ versus $0.91 \pm 0.002$ for the original model. Despite a slight decrease in sensitivity (TPR), the re-trained model maintains high completeness (TPR = 0.85) at a low contamination rate (FPR = 0.2), making it a time-efficient solution with reduced training costs for future automated disturbance-classification tasks. The SSL model is biased towards recall (0.80) rather than precision (0.58), focusing on identifying as many disturbed systems as possible.
\item We find that the smoothness parameter is less of a distinguisher than the asymmetry parameter in detecting merging galaxies. CAS parameters with the original threshold ($A>0.35$) are more conservative, with higher precision (0.77) but lower recall (0.20). This indicates that CAS identifies a relatively clean sample of disturbed or merging systems under the adopted benchmark, but misses a large fraction of the visually disturbed population when compared with visual classification and the SSL model.
\item Using the threshold derived from ROC analysis (\(A > 0.14\)), the CAS method achieves a more balanced performance, with recall increasing substantially from 0.20 to 0.59, albeit with a reduction in precision from 0.77 to 0.48. Under this threshold, however, CAS still performs worse than the SSL model across all evaluated metrics.
\item Both the visual classification and SSL model show a strong positive correlation with stellar mass at the surface brightness depth of our mock images, with the strongest correlation observed in our visual classification. In contrast, the CAS parameters do not exhibit a statistically significant correlation.

\end{itemize}

In conclusion, the comparison of different methods for identifying interaction-driven disturbance signatures provides valuable insights into their strengths and limitations, offering guidance for future automated disturbed galaxy classification tasks. In particular, the SSL-based approach provides a strong balance between transferability and performance, while the classical CAS method remains useful as a conservative selection tool but is much less complete for recovering the broader disturbed population visible in deep imaging.

\section*{Acknowledgements} \label{sec:6}

This research includes computations using the computational cluster Katana supported by Research Technology Services at UNSW Sydney. SB acknowledges funding support from the Australian Research Council through a Discovery Project DP190101943. TNG-100 was run on the HazelHen Cray XC40 system at the High Performance Computing Center Stuttgart as part of project GCS-ILLU of the Gauss Centre for Supercomputing (GCS). Ancillary and test runs of the IllustrisTNG project were also run on the Stampede supercomputer at TACC/XSEDE (allocation AST140063), at the Hydra and Draco supercomputers at the Max Planck Computing and Data Facility, and on the MIT/Harvard computing facilities supported by FAS and MIT MKI. 

\section*{Data Availability} \label{sec:7}
We will make this data available on request.



\bibliographystyle{mnras}
\bibliography{example} 








\bsp	
\label{lastpage}
\end{document}